\documentclass[twocolumn,journal,10pt]{IEEEtran}

\usepackage{amsfonts}
\usepackage{cite}
\usepackage{amsmath,amssymb}
\usepackage[thmmarks,amsmath]{ntheorem}
\usepackage{multicol}
\usepackage{stfloats}
\usepackage{latexsym}
\usepackage{psfrag}
\usepackage{graphicx,subfigure}
\usepackage{graphicx}
\usepackage{mathrsfs}
\usepackage{bbm}
\usepackage{enumerate}
\usepackage{color}
\usepackage{textcomp}

\usepackage{nomencl}
\usepackage{slashbox}
\usepackage{url}
\usepackage{tabularx}
\usepackage{indentfirst}
\usepackage{flushend}
\usepackage{booktabs}  
\usepackage{threeparttable} 
\usepackage{multirow}
\usepackage{etoolbox}
\usepackage{makecell}
\usepackage{epstopdf}
\renewcommand\nomgroup[1]{%
	\item[\bfseries
	\ifstrequal{#1}{A}{Variables}{%
		\ifstrequal{#1}{C}{Sets}{%
			\ifstrequal{#1}{B}{Parameters}{%
			\ifstrequal{#1}{D}{Other Symbols}{}}}}%
	]}

 
\newtheorem{theo}{Theorem}

\newtheorem{lem}{Lemma}

\newtheorem{defi}{Definition}
\newtheorem{rem}{Remark}

\usepackage{algorithm}  
\usepackage{algpseudocode}  
\usepackage{amsmath}  

\makenomenclature
\begin{document}

\title{ Resilient Controller Synthesis Against DoS Attacks for Vehicular Platooning in Spatial Domain 
	\thanks{ }
	\author{Jian Gong, Carlos Murguia, Anggera Bayuwindra and Jinde Cao,~\IEEEmembership{Fellow, IEEE}
	\thanks{Jian Gong is with the Intelligent Transportation System Research Center, Southeast University, Nanjing, 210096, China (e-mail: gongjian\_reus@163.com).}
	\thanks{Carlos Murguia is with the Department of Mechanical Engineering, Eindhoven
		University of Technology, The Netherlands. (Email: c.g.murguia@tue.nl)}
	\thanks{Anggera Bayuwindra is with the School of Electrical Engineering and Informatics, Bandung Institute of Technology, Bandung 40132, Indonesia (e-mail: bayuwindra@itb.ac.id).}
	\thanks{Jinde Cao is with the School of Mathematics, Southeast University, Nanjing 210096, China, and Yonsei Frontier Lab, Yonsei University, Seoul 03722, South Korea (e-mail: jdcao@seu.edu.cn).}
}}
\maketitle
\begin{abstract} 
This paper proposes a vehicular platoon control approach under Denial-of-Service (DoS) attacks and external disturbances.
DoS attacks increase the service time on the communication
network and cause additional transmission delays, which consequently increase the risk of rear-end collisions of vehicles in the platoon.
To counter DoS attacks, we propose a resilient control scheme that exploits polytopic overapproximations of the closed-loop dynamics under DoS attacks. This scheme allows synthesizing robust controllers that guarantee tracking of both the desired spacing policy and spatially varying reference velocity for all space-varying DoS attacks satisfying a hard upper bound on the attack duration. 
In addition, $\mathcal{L}_2$ string stability conditions are derived to ensure that external perturbations do not grow as they propagate through the platoon, thus ensuring the string stability.
Numerical simulations illustrate the effectiveness of the proposed control method.
\end{abstract}	 

\begin{IEEEkeywords}
Vehicular platoon, string stability, DoS attacks, spatial domain
\end{IEEEkeywords}


   \markboth{\
	\ \ \ \ \ \ \ \ \ \ \ \ \ \ \ \ \ \ \ \ \ \ \ \ \ \ \ \ \ \ \ \ \ \
} {}

\section{Introduction}
Increasing traffic demands pose enormous burdens on the existing transportation infrastructure, which lead to severe traffic congestion and casualties.
In recent decades, the development of Intelligent Transportation Systems (ITS) technologies provides practical ways to address these issues.
Cooperative control of vehicular platoons,  as one of the significant applications of ITS, has tremendous potential to improve traffic throughput, driving safety, and fuel economy, which attracts extensive attention from researchers\cite{zheng2017platooning}.

Platoon control aims to ensure that a string of vehicles travel together with a harmonized velocity and a small inter-vehicle gap, which
leads to an increasing road capacity and a decrease in aerodynamic drag \cite{li2018nonlinear}.
Many issues on vehicular platooning have been explored in the existing literature, such as dynamics modeling \cite{xiao2011practical}, choice of spacing
policies \cite{wijnbergen2021nonlinear} and communication topologies \cite{zheng2015stability}, the effect of imperfect communication \cite{gong2018sampled}, and string stability \cite{feng2019string}.
Moreover, advanced control approaches, e.g., model predictive control \cite{hu2022fuel}, sliding mode control \cite{gao2018distributed}, robust control \cite{wang2022real}, and optimal control \cite{wang2022optimal}, have been developed and applied for better control performance.

The spacing policy employed in vehicular platoon systems has a critical impact on car-following behavior, stability, and traffic performance.
The constant spacing policy (CSP) and constant headway policy (CHP) are the most commonly considered policies in vehicular platoon systems \cite{bayuwindra2019extended}, \cite{bayuwindra2019combined}.
The CSP requires a fixed inter-vehicle distance between successive vehicles, potentially leading to higher traffic throughput.
However, its practical application is limited as it constraints the velocity and acceleration of vehicles and thus makes it difficult to deal with varying car-following situations \cite{zheng2022development}.
Moreover, the CSP can only achieve string stability through a leader-following communication topology.
On the other hand, the CHP regulates the desired inter-vehicle gap in terms of the velocity of the following vehicle. 
The CHP can enhance string stability, albeit at the expense of decreased traffic throughput.
Notice that the above spacing policies are adopted under a connotative assumption that the platoon tracks a constant reference velocity of the lead vehicle in the time domain \cite{besselink2017string}.
In contrast, in many practical situations, such as when a platoon travels on a hill road with varying gradients, it is preferable to track varying reference velocity profiles.
In this case, the limited engine power of the following vehicles may lead to unfulfilled platoon behavior as they fail to perfectly track the reference velocity in the time domain \cite{alam2015heavy}. 
In this paper, we adopt a delayed-based spacing policy that ensures that all vehicles in the platoon track varying velocity profiles in the spatial domain \cite{besselink2017string}.

Internal stability and string stability are two fundamental aspects of vehicular platooning.
Internal stability refers to the convergence of the platoon to a desired equilibrium state in the absence of disturbances.
String stability characterizes the propagation of perturbations through out the platoon.
Some existing studies focus on achieving the so-called strong frequency domain string stability (SFSS)\cite{gong2018sampled},\cite{zheng2022development},\cite{naus2010string},\cite{oncu2014cooperative}, relying on frequency domain techniques.
Another commonly used definition is the $\mathcal{L}_p$ string stability (LPSS) proposed in \cite{ploeg2013lp}, intending to regulate the boundedness of system outputs instead of convergence.
In most of these references, e.g.,\cite{gong2018sampled},\cite{zheng2022development},\cite{ploeg2013lp},\cite{di2015design}, the analysis of string stability is limited to given predefined controllers in terms of sufficient conditions for string stability.
Therefore, a synthesis framework that allows the design of platooning controllers that enforce string stability in the presence of disturbances is of high practical and scientific value. 

Another aspect to consider in vehicular platooning is the potential presence of cyber attacks enabled by the cyber-physical nature of modern vehicles and infrastructure.
Several types of cyber attacks to vehicular systems are reported in the literature, e.g., Denial of Service (DoS) attacks, replay attacks, and false data injection attacks \cite{sun2021survey}. 
Among these attack strategies, DoS attacks are the easiest to implement (as channel jamming is enough to induce denial of service) and hence they are one of the most frequent and fatal attacks to vehicular communication networks \cite{he2017survey}. 
DoS attacks maliciously interfere with the information transmission between vehicles.
This could disturb the dynamics of vehicular platoon systems, consequently leading to performance degradation and even vehicle collisions \cite{zhao2021resilient}.
Various security methods have been investigated for vehicular platoon systems to cope with DoS attacks.
However, most existing results focus on exploring the detection mechanism of DoS attacks and evaluating the performance of platoon systems under attacks and given  predefined controllers (e.g., \cite{biron2018real} and \cite{wang2020real}).
As far as the authors are aware, few studies on vehicular platoon control synthesis subject to DoS attacks have been explored in the literature.
In \cite{zhao2021resilient}, by introducing a recovery mechanism to constrain the adverse effects of the DoS attacks, a resilient control protocol is proposed to achieve internal stability of platoon systems.
In \cite{zhang2020distributed}, DoS attacks are modeled as a variety of switching dynamics at sampling instants, and a delay-based platoon control scheme is proposed.
In \cite{ge2022resilient} and \cite{xiao2021secure}, to counter intermittent DoS attacks, resilient controller design methods are presented to achieve secure platooning.
However, the aforementioned studies lose sight of guaranteeing string stability.
In \cite{merco2020hybrid}, by modeling DoS attacks as packet losses, hybird controllers are designed to resist DoS attacks in vehicular platoon and enforce string stability. However, the controller gains are only selected by checking the feasibility of the string stability conditions and fixed to find the upper bound of tolerated DoS attacks using a gridding search method, which is conservative and computationally expensive.
Hence, given the few control design methods available in the literature, it remains attractive to develop resilient controller synthesis methods that are maximally robust against DoS attacks and achieve internal and string stability. This motivation forms the basis of this paper.

In this paper, we study the distributed vehicular platoon control problem in the presence of DoS attacks and external disturbances in the spatial domain. 
Our main contribution is the development of a control synthesis framework to design resilient controllers that ensure internal and string stability of the platoon in spatial domain for peak-bounded disturbances and stochastic DoS attacks which duration satisfy a hard upper bound. We seek to design controllers that maximize the time DoS attacks could be active while maintaining stability (internal and string) of the compromised closed-loop system maximizing thus robustness/resilience against DoS attacks. Under this framework, DoS attacks postpone the information reception of following vehicles, which is modeled as stochastic space-varying delays under the spatial sampling mechanism. To enable the robust controller synthesis, DoS attacks modeled as unknown but bounded space-varying delays are addressed using polytopic overapproximation techniques of the delayed sampled closed-loop dynamics.  The maximal allowable DoS duration and internal and string stability of the closed-loop system are characterized in terms of a set of sufficient conditions posed as LMI constraints by solving a set of LMI conditions.

The structure of this paper is as follows. In Section II, the problem formulation is presented, along with a detailed description of the vehicular platoon model and the objectives of this study. Section III elaborates on the distributed controller design method proposed in this paper, encompassing the polytopic overapproximation technique and controller synthesis approach, which ensure both internal stability and string stability of the platoon system. Section IV presents the numerical simulations, followed by the concluding remarks in Section V.


\begin{table}\renewcommand{\arraystretch}{1.35}
	\caption{Notation}\centering
	\label{notation}
	\begin{tabular}{ll}
		\toprule
		\hline
		Mathematical Notation & Description	\\
		\hline
		$\mathbb{R}^{p\times q}$   &$p\times q$ real matrix set    \\
		$\mathcal{X} >0$   &Positive definite matrix $\mathcal{X}$\\
		$\mathcal{X}^T$   &Transpose of matrix $\mathcal{X}$   \\
		$\text{diag}\{\cdots\}$   &Block-diagonal matrix   \\
		$I$   &Identity matrix    \\
		$0$   &Zero matrix with appropriate dimensions \\
		$\star$   &Symmetric elements of a symmetric matrix \\
		$||z||_{\mathcal{L}_2}$   &$\mathcal{L}_2$ norm of signal $z$, $||z||_{\mathcal{L}_2}=(\int_{0}^{\infty}||z||^2)^{\frac{1}{2}}$\\
		\hline
		System Notation  & Description	\\
		\hline
		$\mathcal{V}_N$   &Set of follower vehicles    \\
		$\mathcal{V}_N^0$   &Set of vehicles including the leading vehicle\\
		$t_i(s)$   &Time instance at which vehicle $i$ pass $s$\\
		$\Delta T$   &Nominal time gap\\
		$\varGamma_i(s)$   &Deviation from $\Delta T$ relative to vehicle $i-1$\\
		$\varGamma_i^0(s)$   &Deviation from $i\Delta T$ relative to the lead vehicle\\
		$v_i(s)$   &Velocity of vehicle $i$ at space $s$\\
		$a_i(s)$   &Acceleration of vehicle $i$ at space $s$\\
		$d_i(s)$   &External disturbance of vehicle $i$ at space $s$\\
		$u_i(s)$   &Control input of vehicle $i$ at space $s$\\
		$\hat{u}_i(s)$, $\bar{u}_i(s)$   & Virtual control input of vehicle $i$\\
		$\zeta$    &Inertial time constant\\
		$v_{\text{ref}}(s)$    &Spatially varying reference velocity\\
		$\delta_{1,i}(s)$    &Velocity tracking error of vehicle $i$\\
		$\delta_{2,i}$    &Space derivative of $\delta_{1,i}(s)$\\
		$\mathcal{E}_{1,i}$    &Time gap tracking error\\
		$\mathcal{E}_{2,i}$    &Space derivative of $\mathcal{E}_{1,i}$\\
		$\varepsilon_0, \varepsilon$    &Weights\\
		$y_i(s)$    &Output of vehicle $i$\\
		$k_1, k_2, k_3$    &Controller gains\\
		$h$    &Space sampling interval\\
		$\tau_{i,k}$    &Transmission delays at sampling instant k\\
		$p$    & Maximum integer multiples of $h$ in $\tau_{i,k}$\\
		$\bar{\tau}_{i,k}$    &Reminder of $\tau_{i,k}$\\
		$\tau_{\text{min}},\tau_{\text{max}}$    &Lower and upper bound of $\bar{\tau}_{i,k}$\\
		$\gamma$    &Upper bound of allowed disturbance propagation\\
		\hline
		\hline
	\end{tabular}
\end{table}

\begin{figure}[t!]
	\centering\includegraphics[height=3.6cm, width=9cm]{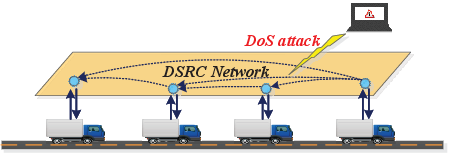}
	\centering\caption{A scenario of a vehicular platoon driving on a straight lane subject to DoS attacks. \label{fig1}}
\end{figure}
\section{Problem Formulation}
Let $\mathcal{V}_N:=\{1, 2, \dots, N\}$ denote the set of follower vehicles, and  $\mathcal{V}_N^0:=\{0, 1, 2, \dots, N\}$ denotes the set of vehicles including the lead vehicle indexed by $0$.
Each vehicle is equipped with a global positioning system and is capable of exchanging state information through the Dedicated Short Range Communication (DSRC) network. 
To exchange information among vehicles, the \textit{ predecessor-leader following topology} is adopted (see Fig. 1), which has significant benefits in ensuring string stability of the platoon \cite{gong2018sampled}.
Additionally, the communication network is assumed to suffer unknown DoS attacks, which interfere with the transmission of real-time vehicle information.
\subsection{Delay-based Spacing Policy}
The choice of spacing policy plays a crucial role in determining the dynamic behavior of the vehicular platoon. The two most commonly used spacing policies are the CSP and CHP \cite{zheng2015stability}. 
However, the CSP and CHP may lead to unsatisfactory platoon behavior as they fail to perfectly track varying reference velocity profiles of the lead vehicle in the time domain due to the limited engine power \cite{alam2015heavy}.
Therefore, in this study, we consider a delay-based spacing policy which ensures that vehicles track the same (varying) velocity profile in space. This policy has been previously studied in \cite{besselink2017string} and is adopted in our research.

Let $s_i(t)$ and $v_i(t)$ denote the longitudinal position and velocity of vehicle $i$, respectively, which satisfies the following kinematic relation
\begin{align}
\dot{s_i}(t):=\frac{ds_i}{dt}=v_i(t).
\end{align}
The delay-based spacing policy describes the desired behavior $s_{\text{des},i}$ of vehicle $i$, and is given by
\begin{align}
s_{\text{des},i}(t)=s_{i-1}(t-\Delta T), \label{timespacingpolicy}
\end{align}
with $i \in \mathcal{V}_N$, where vehicle $i$ tracks the time-delayed position trajectory of its preceding vehicle with time gap $\Delta T >0$.

Assuming that the velocities of all vehicles are always positive, the delay-based spacing policy \eqref{timespacingpolicy} can be equivalently represented in the spatial domain.
Fig. 2 shows an example of time-space trajectories of the lead vehicle, vehicle $i-1$, and vehicle $i$.
Let $t_i(s)$ denote the time instance at which vehicle $i$ pass $s$, where space $s$ is the independent variable.
The spacing policy \eqref{timespacingpolicy} can be expressed as $\varGamma_i(s)=0$, where $\varGamma_i$ represents the deviation from the nominal time gap $\Delta T$ as
\begin{align}
\varGamma_i(s)=t_i(s)-t_{i-1}(s)-\Delta T,\label{spacingpolicy}\\
\varGamma_i^0(s)=t_i(s)-t_0(s)-i\Delta T \label{spacingpolicylead}
\end{align}
with $i \in \mathcal{V}_N$. Likewise, $\varGamma_i^0$ denotes the deviation from the nominal time gap with respect to the lead vehicle of the platoon.
\begin{rem}\label{remark1}
The delay-based spacing policy in \eqref{timespacingpolicy} is transformed into a spatial form in \eqref{spacingpolicy}.
It is worth noting that the spatial spacing policy in \eqref{spacingpolicy} does not require the analysis of time-delay systems as suggested by \eqref{timespacingpolicy}, which facilitates the controller synthesis in the spatial domain.
\end{rem} 
\begin{figure}
	\centering\includegraphics[height=5cm, width=8cm]{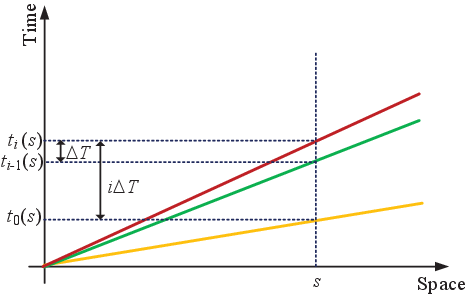}
	\centering\caption{Time-space trajectories of the lead vehicle (yellow), vehicle $i-1$ (green), and vehicle $i$ (red) in the platoon. Time-coordinate indicates time $t$(s) as a function of space $s$. \label{fig2}}
\end{figure}
\subsection{Vehicle Longitudinal Dynamics}
The nonlinear longitudinal vehicle dynamics usually contain the powertrain, longitudinal tire forces, rolling resistance forces, gravitational forces, and aerodynamic drag forces.
Meanwhile, it also embeds uncertainties that involve modeling errors and time-varying external disturbances.
For the convenience of analysis, using the input-output linearization technique \cite{gong2018sampled}, the generalized linear longitudinal dynamics of vehicle $i$ can be described as
\begin{align}
\begin{aligned}
&\dot{s}_i(t)=f(\theta_i(t)),\\
&\dot{\theta}_i(t)=g(\theta_i(t))+h(\theta_i(t))u_i(t)+l(\theta_i(t))d_i(t),
\end{aligned}
\end{align}
with $i \in \mathcal{V}_N^0$, where the state $\theta_i(t)$ denotes a general description of the remaining dynamics, $f(\theta_i(t))$, $g(\theta_i(t))$, $h(\theta_i(t))$, and $l(\theta_i(t))$ represent smooth functions with respect to $\theta_i(t)$, $u_i(t)$ is the control input to be designed and
$d_i(t)$ denotes the unmeasurable external disturbance, which could be also resulting from modeling errors or parameter uncertainties.
By letting $f(\theta_i(t))=v_i(t)$ with $\theta_i=[v_i~a_i]^T$,  the following vehicle dynamics in time domain is adopted \cite{gong2020sampling}
\begin{align}
\begin{aligned}
&\dot{s_i}(t)=v_i(t),
\\&\dot{v_i}(t)=a_i(t)+d_i(t),
\\&\dot{a_i}(t)=-\frac{1}{\zeta}a_i(t)+\frac{1}{\zeta} u_i(t),
\end{aligned} \label{timevehicledynamics}
\end{align}
where $a_i(t)$ denotes the acceleration of vehicle $i$, and $\zeta >0$ is the inertial time constant (considering homogeneous vehicles).

In order to satisfy the spatial spacing policy \eqref{spacingpolicy}, the vehicle dynamics is expected to be written in spatial domain.
By exploiting the kinematic relation (1), the vehicle dynamics in the time domain  \eqref{timevehicledynamics} can be reformulated in space as
\begin{align}
\begin{aligned}
&\dot{t_i}(s)=\frac{1}{v_i(s)},
\\&\dot{v_i}(s)=\frac{1}{v_i(s)}a_i(s)+\frac{1}{v_i(s)}d_i(s),
\\&\dot{a_i}(s)=-\frac{1}{\zeta v_i(s)}a_i(s)+\frac{1}{\zeta v_i(s)} u_i(s),
\end{aligned}
\label{spacevecledynamic}
\end{align}
where space $s \in \mathbb{R}$ serves as independent variable of the system dynamics, and the velocity $v_i(s)$ is assumed to be always positive for all $s\geq 0$.

Note that a nonlinear vehicle dynamics in spatial domain is obtained, which will be further proceeded for platoon modeling using the input-output linearization technique.

\subsection{Longitudinal Platoon Modeling}
Vehicles in the platoon are required to track a prescribed reference velocity and follow the delay-based spacing policy in the spatial domain.
To enable platoon modeling, the vehicle dynamics \eqref{spacevecledynamic} with spacing policy \eqref{spacingpolicy} will be expressed in time gap tracking error coordinates.

Denote the spatially varying reference velocity as $v_{\text{ref}}(s)$, and define the velocity tracking error $\delta_{1,i}$ as well as its space derivative $\delta_{2,i}$
\begin{align}
&\delta_{1,i}(s):=\frac{1}{v_i(s)}-\frac{1}{v_{\text{ref}}(s)},\label{delta1}\\
&\delta_{2,i}(s):=\frac{d }{d s}\delta_{1,i}(s)
\label{e2}
\end{align}
for any following vehicle $i\in \mathcal{V}_N$. Without loss of generality, the function of the reference velocity $v_{\text{ref}}(s)$ is assumed to be continuously differentiable for all $s>0$ and to satisfy $v_{\text{min}}\leq v_{\text{ref}}(s)\leq v_{\text{max}}$ for positive constants $v_{\text{min}}$, $v_{\text{max}}$.

The following controller 
\begin{align}
\begin{aligned}
u_i(s):=&a_i(s)+3\zeta\frac{a_i^2(s)}{v_i(s)}\\
&-\zeta v_i^4(s)\left(\frac{\text{d}^2}{\text{d}s^2}\left(\frac{1}{v_{\text{ref}}(s)}\right)+\hat{u}_i(s)\right) 
\end{aligned}\label{controller}
\end{align}
achieves input\text{-}output linearization of (\ref{spacevecledynamic}) with new virtual input $\hat{u}_i(s)$, such that the dynamics \eqref{spacevecledynamic} with \eqref{controller} can be rewritten as
\begin{align}
\begin{aligned}
&\dot{t_i}(s)=\delta_{1,i}(s)+\frac{1}{v_\text{ref}(s)},
\\&\dot{\delta_i}(s)=A\delta_i(s)+B\hat{u}_i(s)+\xi(\theta_i)d_i(s)
\end{aligned}
\label{vehicledynamics}
\end{align}
for any vehicle $i \in \mathcal{V}_N^0$, where the linear dynamics for $\delta_i:=[\delta_{1,i}, \delta_{2,i}]^{\mathrm{T}}$ is characterized by the matrices
\begin{align}
A=\begin{bmatrix} 0 & 1 \\ 0 & 0 \end{bmatrix}, \ B=\begin{bmatrix} 0  \\ 1  \end{bmatrix}\label{AB}. \notag
\end{align}

 For the case $\theta_i(s)=[v_i(s),a_i(s)]^T$, the disturbance $d_i$ interferes the dynamics \eqref{vehicledynamics} through the function $\xi(\theta_i)=[\xi_1^T(\theta_i),\xi_2^T(\theta_i)]^T$ given by
\begin{align}
\xi_1(\theta_i(s))=-\frac{1}{v_i^3(s)},~ \xi_2(\theta_i(s))=-\frac{3a_i(s)}{v_i^5(s)}.\notag
\end{align}

Based on the velocity tracking error $\delta_{1,i}$ and the spatial spacing policy \eqref{spacingpolicy}, a weighted combination of the time gap tracking error is defined as
\begin{equation}
\mathcal{E}_{1,i}:=(1-\varepsilon_0)\Gamma_i(s)+\varepsilon_0\Gamma_i^0(s)+\varepsilon \delta_{1,i}(s) \label{E1}
\end{equation}
for vehicle $i \in \mathcal{V}_N$ with weights $0\leq \varepsilon_0<1$ and $\varepsilon>0$, where $\Gamma_i$ and $\Gamma_i^0$ are defined in \eqref{spacingpolicy} and \eqref{spacingpolicylead}, respectively. 
The weight $\varepsilon_0$ is selected for penalizing the spacing errors with respect to the preceding vehicle and lead vehicle in the platoon.
Besides, the additional term $\varepsilon \delta_{1,i}$ is introduced to release the spacing policy condition for tracking the desired reference velocity $v_{\text{ref}}$, and to ensure damping of perturbations similar to the case of the CHP strategy.

Using the fact $\dot{\Gamma}_i=\delta_{1,i}-\delta_{1,i-1}$ and $\Gamma_i^0=\Gamma_i+\Gamma_{i-1}^0$, equation \eqref{E1} induces the following dynamics
\begin{equation}
\varepsilon\dot{\Gamma}_i(s)=-\Gamma_i(s)+\mathcal{E}_{1,i}-\varepsilon_0\Gamma_{i-1}^0(s)-\varepsilon \delta_{1,i-1}(s).\\
\label{Gammadynamics}
\end{equation}

The terms with respect to the preceding vehicle $i-1$ in \eqref{Gammadynamics} are collected and defined as
\begin{align}
y_{i-1}(s):=-\varepsilon_0\Gamma_{i-1}^0(s)-\varepsilon \delta_{1,i-1}(s).
\label{output}
\end{align}

Let $\mathcal{E}_{2,i}:=\dot{\mathcal{E}}_{1,i}$ be the additional time gap tracking error coordinate, then 
\begin{equation}
\mathcal{E}_{2,i}=(1-\varepsilon_0)(\delta_{1,i}-\delta_{1,i-1})+\varepsilon_0(\delta_{1,i}-\delta_{1,0})+\varepsilon \delta_{2,i}(s).\label{E2}
\end{equation}

Now, the platoon dynamics can be written using the timing error coordinates $x_i=[\Gamma_i,\mathcal{E}_{1,i},\mathcal{E}_{2,i}]^T$, where $\Gamma_i$ represents the desired spacing policy in \eqref{spacingpolicy}.
Here, introducing a new virtual input $\bar{u}_i$ by substituting 
\begin{equation}
\hat{u}_i(s)=-\varepsilon^{-1}(1-\varepsilon_0)(\delta_{2,i}-\delta_{2,i-1})-\\
\varepsilon^{-1}\varepsilon_0(\delta_{2,i}-\delta_{2,0})+\bar{u}_i(s)
\label{uF}
\end{equation}
into \eqref{vehicledynamics}, one has a cascaded system of the platoon dynamics in timing error coordinates $x_i$ as
\begin{align}
\begin{aligned}
&\dot{x_i}(s)=\bar{f}(x_i(s),\bar{u}_i(s),y_{i-1}(s),\bar{d}_i(s)), 
\\&y_i(s)=\bar{h}(x_i(s)),
\end{aligned}
\label{platoondynamic}
\end{align}
where $y_i(s)$ denotes the output characterized by \eqref{output}. 
By recalling the dynamics for $\Gamma_i(s)$ in \eqref{Gammadynamics} and the definitions \eqref{E1} and \eqref{E2}, the vector field $\bar{f}$ is given by
\begin{align}
\bar{f}(x_i,\bar{u}_i,y_{i-1},\bar{d}_i)=\begin{bmatrix} -\varepsilon^{-1}(-\Gamma_i+\mathcal{E}_{1,i}+y_{i-1})  \\ A\mathcal{E}_i+ \varepsilon B\bar{u}_i+\bar{d}_i \end{bmatrix}\ \notag
\end{align}
with $\mathcal{E}_i=[\mathcal{E}_{1,i},\mathcal{E}_{2,i}]^T$. Using \eqref{E1} and \eqref{output}, it yields the output equation
\begin{align}
\bar{h}(x_i)=(1-\varepsilon)\Gamma_i-\mathcal{E}_{1,i}.
\end{align}
Moreover, given by \eqref{vehicledynamics} and \eqref{platoondynamic},  the disturbance $\bar{d}_i$ takes the following form
 \begin{equation}
 \bar{d}_i(s)=\bar{\xi}(\theta_i, \theta_{i-1}, \theta_0)\tilde{d}_i(s)
 \end{equation}
with $\tilde{d}_i(s)=[d_i(s), d_{i-1}(s), d_0(s)]^{\mathrm{T}}$ and $\bar{\xi}(\theta_i, \theta_{i-1}, \theta_0)=[ \bar{\xi}_1^{\mathrm{T}}(\theta_i, \theta_{i-1}, \theta_0), \bar{\xi}_2^{\mathrm{T}}(\theta_i, \theta_{i-1}, \theta_0)]^{\mathrm{T}}$, where
\begin{equation}
\bar{\xi}_1(\theta_i, \theta_{i-1}, \theta_0)=\begin{bmatrix} \varepsilon \xi_1^{\mathrm{T}}(\theta_i) \\ 0 \\ 0   \end{bmatrix} ^{\mathrm{T}},\notag
\end{equation}
\begin{equation}
\bar{\xi}_2(\theta_i, \theta_{i-1}, \theta_0)=\begin{bmatrix} \varepsilon \xi_2^{\mathrm{T}}(\theta_i)+\xi_1^{\mathrm{T}}(\theta_i) \\ (\varepsilon_0-1)\xi_1^{\mathrm{T}}(\theta_{i-1}) \\ -\varepsilon_0 \xi_1^{\mathrm{T}}(\theta_0)   \end{bmatrix} ^{\mathrm{T}}. \notag
\end{equation}
Here, one can observe that the disturbances on both the preceding vehicle and lead vehicle influence the time gap error of vehicle $i$.

Overall, the platoon dynamics can be written in the state space form as
\begin{align}
\dot{x}_i=\bar{A}_0x_i+\bar{B}_1\bar{u}_i+\bar{B}_2y_{i-1}+\bar{B}_3\bar{d}_i \label{platoonstatedynamics}
\end{align}
with $x_i=[\Gamma_i,\mathcal{E}_{1,i},\mathcal{E}_{2,i}]^T$,
\begin{align}
&\bar{A}_0=\begin{bmatrix} -\varepsilon^{-1} &\varepsilon^{-1} &0  \\ 0 &0 &1 \\
0 &0 &0 \end{bmatrix}, \bar{B}_1=\begin{bmatrix} 0 \\0 \\\varepsilon \end{bmatrix},
\bar{B}_2=\begin{bmatrix} \varepsilon^{-1}  \\ 0 \\
0 \end{bmatrix},\notag\\
&\bar{B}_3=\begin{bmatrix} 0 &0 \\1 &0  \\ 0 &1  \end{bmatrix},\notag
\end{align}
where $y_{i-1}$ represents the external input from the preceding vehicle $i-1$, and $\bar{d}_i$ denotes the external disturbance input of vehicle $i$.

Moreover, the output equation is given by
\begin{align}
y_i=Cx_i
\end{align}
with $C=[1-\varepsilon,-1,0]^{\mathrm{T}}$.

A distributed state feedback controller can be implemented for the new input $\bar{u}_i$ in \eqref{platoonstatedynamics} as
\begin{align}
\bar{u}_i=-Kx_i, \label{controllaw}
\end{align}
where $K=[k_1,k_2,k_3]$ denotes the controller gains to be designed.

\begin{rem}
Note that the time gap errors $\Gamma_i$ and $\Gamma_i^0$ in \eqref{spacingpolicy} and \eqref{spacingpolicylead}, respectively, are not defined for the lead vehicle (index $i=0$). 
Instead, let $\Gamma_0:=t_0-\int v_{\text{ref}}^{-1}$ and $\Gamma_0^0:=\Gamma_0$
be the deviation from a nominal trajectory.
According to definitions of \eqref{E1} and \eqref{E2}, $\mathcal{E}_{1,0}$ and $\mathcal{E}_{2,0}$ are defined as $\mathcal{E}_{1,0}=\Gamma_0+\varepsilon \delta_{1,0}$ and $\mathcal{E}_{2,0}=\delta_{1,0}+\varepsilon \delta_{2,0}$, respectively.
In addition, the disturbance for the lead vehicle in the platoon can be obtained as $\bar{d}_0=\bar{\xi}(0, 0, \theta_0)\tilde{d}_0(s)$, where $\tilde{d}_0(s)=[0, 0 , d_0^\mathrm{T}(s)]^{\mathrm{T}}$. 
Then, one can conclude that the dynamics of the lead vehicle has the same form as that of the following vehicles in \eqref{platoondynamic}.
\end{rem}

\subsection{DoS Attacks Model}
Denial of Service attacks pose a serious threat to cooperative vehicle control in V2V networks. These attacks occur when an attacker inserts fake or irrelevant messages into the communication channel, rendering the network inaccessible to legitimate vehicles \cite{sun2021survey}. As a result, DoS attacks can cause delays in information transmission, leading to additional service time. In the context of vehicular platooning, such delays can significantly increase the risk of collisions between adjacent vehicles. The effect of DoS attacks on service time has been modeled as different types of time delays in the existing literature, such as constant time delays and probabilistic time delays \cite{biron2018real}.

\begin{figure}
	\centering\includegraphics[height=5.2cm,width=8cm]{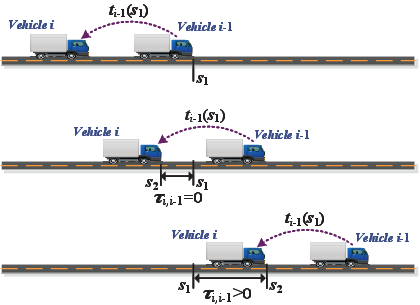}
	\centering\caption{A scenario of vehicle information transmission from vehicle $i-1$ to vehicle $i$ with spatial delays under DoS attacks. Vehicle $i-1$ samples its information at space $s=s_1$, and transmits to vehicle $i$ (top row). Vehicle $i$ receives the sampled information of vehicle $i-1$ at space $s=s_2$, where two cases are included: 1) Case 1: $s_2 \leq s_1$ (middle row); 2) Case 2: $s_2 > s_1$ (bottom row).
		\label{fig5}}
\end{figure}

In this paper, considering the randomness of DoS attacks, the platoon control system is assumed to suffer unknown DoS attacks such that vehicles receive the state information of their preceding vehicle and the lead vehicle with stochastic and space-varying delays. 
In particular, recalling the platoon dynamics \eqref{platoonstatedynamics} and the definitions of $\mathcal{E}_{1,i}$ and $\mathcal{E}_{2,i}$ in \eqref{E1} and \eqref{E2}, the time information of vehicle $i-1$ and the lead vehicle with respect to space $s$, i.e., $t_{i-1}(s)$ and $t_0(s)$, is requested for the control law \eqref{controllaw} of vehicle $i$.

Fig.3 shows examples of vehicle information transmission between vehicle $i$ and its preceding vehicle under DoS attacks. 
Specifically, vehicle $i-1$ samples the state information $t_{i-1}(s_1)$ at space $s=s_1$, and then transmits it to vehicle $i$.
Due to spatial delays induced by DoS attacks, vehicle $i$ receives the state information of its preceding vehicle $t_{i-1}(s_1)$ at space $s=s_2$.
Denote $\tau_{i,i-1}(s_1)$ as the resulting delay in information transmission from the preceding vehicle at space instant $s=s_1$.   
There are two cases to explore the spatial delay in the control system of vehicle $i$: 
1) Case 1: if $s_2\leq s_1$, the delay $\tau_{i,i-1}(s_1)$ affecting the controller of vehicle $i$ is equal to zero. This is because vehicle $i$ has received the information of its preceding vehicle $t_{i-1}(s_1)$ before reaching space $s_1$. Therefore, no delay impacts the controller of vehicle $i$ at space instant $s=s_1$.
2) Case 2: if $s_2> s_1$, the spatial delay for the controller of vehicle $i$ at $s=s_1$ can be computed as $\tau_{i,i-1}(s_1)=s_2-s_1$.
Similar situations can be applied in calculating the spatial delay in the information transmission from the lead vehicle, i.e., $\tau_{i,0}(s_1)$.  

Note that the sampler is space-driven, and samples the state information of each vehicle at sampling instant $s_k$. The constant sampling period is defined as $h=s_{k+1}-s_k$.
Let $\tau_{ij,k}$ for $i \in \mathcal{V}_N$, $j=i-1~ \text{or} ~0$, denote the  spatial delay caused by the information transmission from the preceding vehicle or lead vehicle at sampling instant $s_k$. 
The transmission delay induced by the DoS attacks for vehicle $i$ can be expressed as
the maximum delay in that from the preceding vehicle and lead vehicle, and is given by $\tau_{i,k}:= \max \limits_{j}\{\tau_{ij,k}|j=i-1~ \text{or} ~0\}$.

Moreover, the space-varying and probably large transmission delay $\tau_{i,k}$ for vehicle $i \in \mathcal{V}_N$, can be modeled as
\begin{equation}
\tau_{i,k}=\bar{\tau}_{i,k}+(p-1)h,~~~\bar{\tau}_{i,k}\in[0,h], ~~p\in \{1,2,...,\}, \label{tau_p}
\end{equation}
where the large delays imply to be larger than the sampling interval $h$ for $p>1$. Considering the DoS attacks, the continuous-space closed-loop model for the platoon system can be written as 
\begin{align}
\begin{aligned}
&\dot{x}_i(s)=\bar{A}_0x_i(s)+\bar{B}_1\bar{u}_i(s)+\bar{B}_2y_{i-1}(s)+\bar{B}_3\bar{d}_i(s),\\
&\bar{u}_i(s)=\bar{u}_{i,k-p+1},~~s\in [s_k+\bar{\tau}_{i,k}, s_{k+1}+\bar{\tau}_{i,k}],
\end{aligned}
\label{continousmodel}
\end{align}
where $\bar{u}_{i,k}:=\bar{u}_i(s_k)$, and a zero-order holder (ZOH) transforms the delayed discrete-space control input $\bar{u}_{i,k}$ to the continuous-space control input $\bar{u}_i(s)$.
To capture the effect of DoS attacks modeled as space delays, defining the discrete space signals, i.e., $x_{i,k}:=x_{i}(s_k)$, the continuous-space closed-loop model \eqref{continousmodel} is exactly discretized at $s_k$ as
\begin{align}
x_{i,k+1}=&e^{\bar{A}_0h}x_{i,k}+\int_{0}^{h-\bar{\tau}_{i,k}}e^{\bar{A}_0s}ds\bar{B}_1\bar{u}_{i,k-p+1} \notag
\\&+\int_{h-\bar{\tau}_{i,k}}^{h}e^{\bar{A}_0s}ds\bar{B}_1\bar{u}_{i,k-p}
+\int_{0}^{h}e^{\bar{A}_0s}ds\bar{B}_2y_{i-1,k} \notag
\\ &+\int_{0}^{h}e^{\bar{A}_0s}ds\bar{B}_3\bar{d}_{i,k}. \label{discreteplatoonmodel}
\end{align}

\begin{rem}\label{remark3} 
	In the existing literature, the effects of DoS attacks in vehicular platoon control systems have been mainly formulated by two control-oriented perspectives, i.e., packet dropouts \cite{zhao2021resilient}\cite{merco2020hybrid} and time delays \cite{biron2018real}\cite{biron2017resilient}.
	The former formulation would be inherently limiting and fail to capture the intelligent and targeted nature of DoS attacks due to the random characterization of DoS attacks \cite{de2014resilient}.
	The later modeling method of DoS attacks better formulates the attacker's intelligence since the actions of the attacker are not easily observed by packet loss detection algorithms.
	Therefore, this paper models the effect of DoS attacks on service time using stochastic delays, which provides a more realistic representation of the attack's impact.
\end{rem}
\begin{rem}\label{remark4} 
	In addition, the relative existing methods focus on modeling the effect of DoS attacks in terms of time sampling , which influences the platoon dynamics in the time domain.
	In fact, modeling the effects of DoS attacks in the spatial domain is more intuitive and crucial, as it clearly indicates its impact on vehicle spacing in the platoon.
\end{rem}
\begin{rem}\label{remark5} 
We assume that the attacker has a limited resources to interfere with V2V networks. We capture this by assuming that the unknown delay caused by DoS attacks is upper bounded, i.e., $p \in [0,~ p_{\text{max}}]$, for some positive constant $p_{\text{max}}$.
\end{rem}

\subsection{Control Synthesis Objective}
This paper aims to develop space-based platoon control strategies in the presence of DoS attacks and external disturbances. 
The following definition of $\mathcal{L}_2$ string stability is introduced, which is used below to characterize the effect of disturbances in the platooning dynamics and state the control objectives we seek to address.  

\begin{defi}\cite{ploeg2013lp}
 Consider the following cascaded platoon system
\begin{align}
&\dot{x}_0=\bar{f}(x_0, \bar{u}_0),\notag\\
&\dot{x}_i=\bar{f}(x_i,x_{i-1}),\notag\\
&y_i=\bar{h}(x_i),
 \label{def1}
\end{align}
where $\bar{u}_0$ is the external input of the lead vehicle, $x_i$ is the state of vehicle $i \in \mathcal{V}_N$, and $y_i$ is the output.
Denote $x=[x_0^T,x_1^T,\cdots ,x_N^T]^T$ as the lumped
state vector and $\tilde{x}=[\tilde{x}_0^T,\tilde{x}_0^T,\cdots, \tilde{x}_0^T]^T$ as the constant equilibrium solution of \eqref{def1} for $\bar{u}_0=0$.
The platoon system \eqref{def1} is $\mathcal{L}_2$ string stable if there exist class $\mathcal{K}$ functions $\alpha$ and $\beta$, such that, for any control input for the lead vehicle $\bar{u}_0$ and initial state $x(0)$
\begin{align}
||y_i(s)-\bar{h}(\tilde{x}_0)||_{\mathcal{L}_2} \leq \alpha (||\bar{u}_0||_{\mathcal{L}_2})+\beta (||x(0)-\tilde{x}||),
\end{align}
with $i \in \mathcal{V}_N$. In addition, if $x(0)=\tilde{x}$, it also holds that 
\begin{align}
||y_i(s)-\bar{h}(\tilde{x}_0)||_{\mathcal{L}_2}\leq ||y_{i-1}(s)-\bar{h}(\tilde{x}_0)||_{\mathcal{L}_2}
\end{align}
for $i \in \mathcal{V}_N$, the platoon system \eqref{def1} is strictly $\mathcal{L}_2$ string stable.

\end{defi}

 The control objectives are the following:
\begin{enumerate}
\item[1.] \textit{Internal stability}: It is required that the platoon dynamics \eqref{discreteplatoonmodel} for vehicle $i, i\in \mathcal{V}_N$, converges to the
origin, i.e., $\lim\limits_{s_k \to \infty} x_{i,k}=0$, in the absence of
the external inputs. 
To be specific, in terms of the definition of state $x_{i,k}$, the spacing policy \eqref{spacingpolicy} is satisfied such that $\lim\limits_{s_k \to \infty} \Gamma_i(s_k)=0$, which also leads to $\lim\limits_{s_k \to \infty} \Gamma_i^0(s_k)=0$, and 
each vehicle can track the prescribed trajectory of the reference velocity $v_{\text{ref}}(s_k)$, i.e., $\lim\limits_{s_k \to \infty} \delta_{1,i}(s_k)=0$.
\item[2.] \textit{$\mathcal{L}_2$ string stability}: The outputs of vehicles are not amplified when they are propagated downstream along the platoon for any change of the reference velocity.
According to Definition 1, the platoon system \eqref{discreteplatoonmodel} is strictly $\mathcal{L}_2$ string stable if there exist a positive scalar $0< \sigma \leq 1$ such that the following condition holds
\begin{align}
||y_{i, k}||_{\mathcal{L}_2}\leq \sigma||y_{i-1,k}||_{\mathcal{L}_2}. \label{stringstability}
\end{align}
\item[3.] \textit{Disturbance propagation attenuation}: It is required to attenuate the propagation of external disturbances along the platoon. 
Let $\Theta_i(j\omega)$ denote the frequency response describing the relation between the output $y_i$ and disturbance $\bar{d}_i$ of vehicle $i$. The propagation attenuation of the external disturbance on the output can be characterized by 
\begin{align}
\mathop{\text{sup}}\limits_{\omega}|\Theta_i(j\omega)|=\mathop{\text{sup}}\frac{||y_{i,k}||_{\mathcal{L}_2}}{||\bar{d}_{i,k}||_{\mathcal{L}_2}}\leq \gamma,
\end{align}
for $i\in \mathcal{V}_N$, where $\gamma>0$ is a given bound representing the level of allowed maximum disturbance propagation.
\end{enumerate}

\section{Distributed Platoon Controller Design}

\subsection{Polytopic Overappromaximation}
\begin{figure}
	\centering\includegraphics[height=4cm,width=4.8cm]{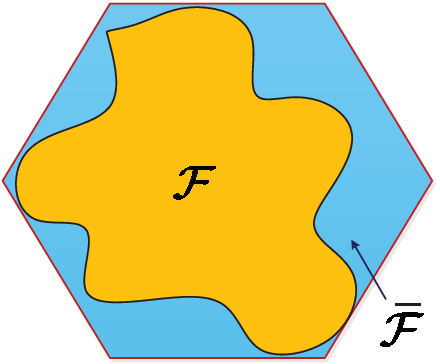}
	\centering\caption{Polytopic overapproximation. \label{polytopic}}
\end{figure}

In the resulting discretized platoon model \eqref{discreteplatoonmodel},  space-varying delays induced by DoS attacks appear in exponential form, which makes it challenging to design a stabilizing controller directly. 
To address this issue, overapproximation techniques are introduced to embed the original model with the exponential uncertainty into a polytope. This will allow us to design robust controllers. 
Available overapproximation methods are based on real Jordan forms, gridding and norm-bounding techniques, Cayley-Hamilton theorem, and Taylor series \cite{heemels2010comparison}. 
The core idea of these overapproximation methods is to construct a polytopic set  $\bar{\mathcal{F}}$, as depicted in Fig. \ref{polytopic}, satisfying 
\begin{align}
\mathcal{F}=\left\{
\int_{0}^{h-\bar{\tau}_{i,k}}e^{\bar{A}_0s}ds|\bar{\tau}_{i,k}\in [\tau_{\text{min}}, \tau_{\text{max}}]
\right\}\subseteq  \bar{\mathcal{F}},
\end{align}
for some polytope vertices  such that the uncertainty set $\mathcal{F}$ is embedded in $\bar{\mathcal{F}}$.
This paper adopts the overapproximation method based on real Jordan forms of the system matrix $\bar{A}_0$ given by
\begin{align}
\mathcal{J}:=\mathcal{Q}^{-1}\bar{A}_0\mathcal{Q},
\end{align}
where $\mathcal{Q}$ is an invertible matrix that consists of the generalized eigenvectors of $\bar{A}_0$, and $\mathcal{J}$ is a block diagonal matrix given by
\begin{align}
\mathcal{J}=\text{diag}(\mathcal{J}_1,..., \mathcal{J}_m),
\end{align}
where $\mathcal{J}_r, r=1,\cdots, m$, denotes the real Jordan block corresponding to  a real eigenvalue $\lambda_r \in \mathbb{R}$ or a pair of complex conjugate eigenvalues $\alpha_r \pm \beta_ri$.

Using the real Jordan decomposition on the exponential integral term $\int_{0}^{h-\bar{\tau}_{i,k}}e^{\bar{A}_0s}ds$ in \eqref{discreteplatoonmodel}, we have 
\begin{align}
\begin{aligned}
\mathcal{F}(\bar{\tau}_{i,k})&=\int_{0}^{h-\bar{\tau}_{i,k}}e^{\bar{A}_0s}ds=\mathcal{Q}^{-1}\int_{0}^{h-\bar{\tau}_{i,k}}e^{\mathcal{J}s}ds\mathcal{Q}\\
&=\mathcal{F}_0+\sum_{n=1}^{\nu}q_n(\bar{\tau}_{i,k})\mathcal{F}_n
\end{aligned}
\end{align}
with matrices $\mathcal{F}_0, \mathcal{F}_1,...,\mathcal{F}_{\nu}$ as introduced in (38), where $\nu$ is the degree of the minimal polynomial \cite{horn2012matrix} of matrix $\bar{A}_0$, and $q_n(\bar{\tau}_{i,k})$ is a function of the space-varying delays that corresponds to the distinct eigenvalues of the matrix $\bar{A}_0$ given by
\begin{align}
\left\{\begin{aligned}
&q_1(\bar{\tau}_{i,k})=h-\bar{\tau}_{i,k},\\
&q_2(\bar{\tau}_{i,k})=(h-\bar{\tau}_{i,k})^2,\\
&q_3(\bar{\tau}_{i,k})=e^{-2(h-\bar{\tau}_{i,k})},
\end{aligned}
\right.
\end{align}
with $n=1,..., \nu, \nu=3$.
\begin{figure*}[t!]
	\begin{align}
	\begin{aligned}
	&\int_{0}^{h-\bar{\tau}_{i,k}}e^{\bar{A}_0s}ds=\mathcal{Q}\int_{0}^{h-\bar{\tau}_{i,k}}
	\begin{bmatrix}
	1 &s &0\\
	0 &1 &0\\
	0 &0 &e^{-2s}
	\end{bmatrix}ds\mathcal{Q}^{-1}\\
	=&\mathcal{Q}
	\begin{bmatrix}
	h-\bar{\tau}_{i,k} &\frac{1}{2}(h-\bar{\tau}_{i,k})^2 &0\\
	0 &h-\bar{\tau}_{i,k} &0\\
	0 &0 &-\frac{1}{2}e^{-2(h-\bar{\tau}_{i,k})}+\frac{1}{2}
	\end{bmatrix}\mathcal{Q}^{-1}\\
	=&\underbrace{\mathcal{Q}
		\begin{bmatrix}
		0 &0 &0\\
		0 &0&0\\
		0 &0 &\frac{1}{2}
		\end{bmatrix}\mathcal{Q}^{-1}}_{\mathcal{F}_0}+
	q_1(\bar{\tau}_{i,k})\underbrace{\mathcal{Q}\begin{bmatrix}
		1 &0 &0\\
		0 &1&0\\
		0 &0 &0
		\end{bmatrix}\mathcal{Q}^{-1}}_{\mathcal{F}_1}+q_2(\bar{\tau}_{i,k})\underbrace{\mathcal{Q}\begin{bmatrix}
		0 &\frac{1}{2} &0\\
		0 &0&0\\
		0 &0 &0
		\end{bmatrix}\mathcal{Q}^{-1}}_{\mathcal{F}_2}+q_3(\bar{\tau}_{i,k})\underbrace{\mathcal{Q}\begin{bmatrix}
		0 &0 &0\\
		0 &0&0\\
		0 &0 &-\frac{1}{2}
		\end{bmatrix}\mathcal{Q}^{-1}}_{\mathcal{F}_3}\label{Overapproaximation}
	\end{aligned}
	\end{align}
\end{figure*}

Lower and upper bounds of the function $q_i(\bar{\tau}_{i,k})$ can thus be defined as 
\begin{align}
\begin{aligned}
&\underline{q}_n:=\min \limits_{\bar{\tau}_{i,k} \in [\tau_{\text{min}}, \tau_{\text{max}}]}q_n(\bar{\tau}_{i,k}),\\
&\overline{q}_n:=\max \limits_{\bar{\tau}_{i,k} \in [\tau_{\text{min}}, \tau_{\text{max}}]}q_n(\bar{\tau}_{i,k}).
\end{aligned}
\label{upperlowerbound}
\end{align}
Defining $\text{Co}\{V_1,..., V_n\}$ as the convex hull of $\{V_1,..., V_n\}$, for $n=1,...,\nu$, we have 
\begin{align}
\mathcal{F}(\bar{\tau}_k^i) \in \text{Co}\{V_1,..., V_{2^\nu}\},
\end{align}
for all $\bar{\tau}_{i,k} \in [\tau_{\text{min}}, \tau_{\text{max}}]$ with
\begin{align}
\{V_1,..., V_{2^\nu}\}:=\left\{\mathcal{F}_0+\sum_{n=1}^{\nu}\eta_n\mathcal{F}_n|\eta_n \in \{\underline{q}_n, \overline{q}_n\}\right\}.\label{41}
\end{align}

From \eqref{41}, we can obtain $2^\nu$ polytope vertices of $\mathcal{F}(\bar{\tau}_k^i)$ and thus construct polytopic set $\bar{\mathcal{F}}$, which will benefit for the robust controller synthesis by solving a finite number of LMIs.

\subsection{Internal Stability}
In this subsection, conditions to handle the effects of DoS attacks are derived so that internal stability of the platoon system in \eqref{discreteplatoonmodel} can be guaranteed in the absence of driving terms from the preceding vehicle and exogenous disturbances. 

Defining the augmented state vector $\mathcal{X}_{i,k}:=[{x_{i,k}}^T, \bar{u}_{i,k-1}^T,\bar{u}_{i,k-2}^T, \cdots,\bar{u}_{i,k-p}^T]^T$, for $y_{i-1, k}=\bar{d}_{i,k}=0$, the discrete-space distributed platoon model \eqref{discreteplatoonmodel} can be rewritten as
\begin{align}
\mathcal{X}_{i,k+1}=\bar{A}(\bar{\tau}_{i,k})\mathcal{X}_{i,k}+\bar{B}(\bar{\tau}_{i,k})\bar{u}_{i,k} \label{augumentplatoonmodel}
\end{align}
with 
\begin{align}
\bar{A}(\bar{\tau}_{i,k})=\begin{bmatrix} 
e^{\bar{A}_0h} &Y_{p-1} &Y_{p-2} &\cdots &Y_0\\
0 &0&0&0&0\\
0 &I &0 &\cdots &0\\
\vdots & &\ddots & \cdots &\vdots\\
0 &\cdots &0 &I &0 \notag
\end{bmatrix},
\end{align}
\begin{align}
\bar{B}(\bar{\tau}_{i,k})=\begin{bmatrix} 
Y_p^T &I &0&\cdots &0 \notag
\end{bmatrix}^T,
\end{align}
\begin{align}
Y_r=
\left\{  
\begin{aligned}
&\int_{h-T_{r+1}}^{h-T_r}e^{\bar{A}_0s}ds\bar{B}_1,~ \text{if}~~ 0 \leq r \leq 1,\\  
&~0, ~~~~~~~~~~~~~~~~~~~~~\text{if}~~ 1 < r \leq p,
\end{aligned} \notag
\right.
\end{align}
where $T_0:=0$, $T_1:=\bar{\tau}_{i,k}$, $T_2:=h$, and $p$ is a positive integer introduced in \eqref{tau_p}.

We consider a structured state feedback control law for the augmented platoon system \eqref{augumentplatoonmodel} as follows
\begin{align}
\bar{u}_{i,k}=-\bar{K}\mathcal{X}_{i,k}, \label{augumentcontrollaw}
\end{align}
where $\bar{K}=[K~~ 0]$ (i.e., we do not feed past inputs back, and only consider the current state), and $K=[k_1,k_2,k_3]$ is the controller gain.

Then, applying the real Jordan form to the exponential terms of the system matrix $\bar{A}_0$ in \eqref{discreteplatoonmodel}, the platoon model \eqref{augumentplatoonmodel} can be written as
\begin{align}
\begin{aligned}
\mathcal{X}_{i,k+1}=&\left(\mathscr{M}_0+\sum_{n=1}^{\nu}q_n(\bar{\tau}_{i,k})\mathscr{M}_n\right)\mathcal{X}_k\\
&+\left(\mathscr{H}_0+\sum_{n=1}^{\nu}q_n(\bar{\tau}_{i,k})\mathscr{H}_n\right)\bar{u}_{i,k},
\end{aligned}\label{polynomialplatoonmodel}
\end{align}
where $\mathscr{M}_n$ and $\mathscr{H}_n$ are the resulted polynomial matrices, and $\nu$ denotes the number of functions $q_n(\bar{\tau}_{i,k})$. Then, we can define the uncertainty sets of the system matrices as 
\begin{align}
\mathscr{M}=\left\{
\mathscr{M}_0+\sum_{n=1}^{\nu}q_n(\bar{\tau}_{i,k})\mathscr{M}_n|\bar{\tau}_{i,k}\in [\tau_{\text{min}}, \tau_{\text{max}}]
\right\},\label{F}
\end{align}
\begin{align}
\mathscr{H}=\left\{
\mathscr{H}_0+\sum_{n=1}^{\nu}q_n(\bar{\tau}_{i,k})\mathscr{H}_n|\bar{\tau}_{i,k}\in [\tau_{\text{min}}, \tau_{\text{max}}]
\right\},\label{G}
\end{align}
where all possible matrix combinations are involved in the infinite dimensional sets $\mathscr{M}$ and $\mathscr{H}$. 
This makes it challenging to analyze stability of system \eqref{polynomialplatoonmodel} directly.
To address this issue, we present a convex overapproximation of the sets $\mathscr{M}$ and $\mathscr{H}$.
 According to the lower and upper bounds of the function $q_i(\bar{\tau}_{i,k})$ in \eqref{upperlowerbound}, overapproximations $\bar{\mathscr{M}}$ and $\bar{\mathscr{H}}$ satisfying $\mathscr{M} \subseteq \bar{\mathscr{M}}$ and $\mathscr{H} \subseteq \bar{\mathscr{H}}$ can be written as
 \begin{align}
 \bar{\mathscr{M}}=\left\{
 \mathscr{M}_0+\sum_{n=1}^{\nu}\eta_n\mathscr{M}_n|\eta_n \in [\underline{q}_n, \overline{q}_n], n=1,..., \nu
 \right\},
 \end{align}
 \begin{align}
 \bar{\mathscr{H}}=\left\{
 \mathscr{H}_0+\sum_{n=1}^{\nu}\eta_n\mathscr{H}_n|\eta_n \in [\underline{q}_n, \overline{q}_n], n=1,..., \nu
 \right\}.
 \end{align}

The sets of vertices $\bar{\mathscr{M}}$ and $\bar{\mathscr{H}}$ can be represented as $\bar{\mathscr{M}}=\left\{ \mathcal{S}_{M,j}|j=1,...,2^{\nu} \right\}$ and $\bar{\mathscr{H}}=\left\{ \mathcal{S}_{H,j}|j=1,...,2^{\nu} \right\}$,
 a finite number of LMI conditions for stabilizing controller design can be posed.

Before deriving internal stability conditions, the following instrumental lemmas are given.

\begin{lem} \cite{cloosterman2009stability}
	Consider the uncertain closed-loop system
	$\mathcal{X}_{i,k+1}=(\bar{A}(\bar{\tau}_{i,k})-\bar{B}(\bar{\tau}_{i,k})\bar{K})\mathcal{X}_{i,k}$ in \eqref{augumentplatoonmodel}. There exists a common quadratic Lyapunov function $V(\mathcal{X}_{i,k})=\mathcal{X}_{i,k}^TP\mathcal{X}_{i,k}$ such that global asymptotic stability of \eqref{augumentplatoonmodel} is guaranteed if the following matrix inequalities are satisfied:
	\begin{align}
	P=P^T&>0, \notag\\
	(\bar{\mathcal{M}}-\bar{\mathcal{H}}\bar{K})^TP(\bar{\mathcal{M}}-\bar{\mathcal{H}}\bar{K})-P&<-\mu P
	\label{lem1}
	\end{align}
\end{lem}
with $\bar{\mathcal{M}} \in \mathscr{M}$ and $\bar{\mathcal{H}} \in \mathscr{H}$ as defined in \eqref{F} and \eqref{G} respectively, and a scalar $0 < \mu <1$.

\begin{lem}
	Consider the discrete-space model of the augmented platoon system in \eqref{augumentplatoonmodel} subject to the space-varying delay with $\bar{\tau}_{i,k} \in [\tau_{\rm min}, \tau_{\rm max}]$. If there exists a $P=P^T>0, P\in \mathbb{R}^{(3+p)\times (3+p)}$ and a scalar $0 < \mu <1$, such that
	\begin{align}
	\begin{bmatrix}
	(1-\mu)P &(\mathcal{S}_{M,j}^T-\bar{K}^T\mathcal{S}_{H,j}^T)P\\
	\star &P
	\end{bmatrix}>0 \label{lemma2}
	\end{align}
	with $ \mathcal{S}_{M,j} \in \bar{\mathscr{M}}$ and $\mathcal{S}_{H,j} \in \bar{\mathscr{H}}$, and $j=\{1,2,...,2^\nu\}$, then the platoon system \eqref{augumentplatoonmodel} is globally asymptotically stable for any delay $\bar{\tau}_{i,k} \in [\tau_{\rm min}, \tau_{\rm max}]$.
\end{lem}

\textbf{\textit{Proof}:} See Appendix A.

Next, based on the control law \eqref{augumentcontrollaw}, a set of LMI conditions are derived in Theorem 1 to guarantee the asymptotic stability of the platoon system \eqref{augumentplatoonmodel}.

\begin{theo}\label{theo1}
	Consider the discrete-space model of the platoon system in \eqref{augumentplatoonmodel} subject to the space-varying transmission delays caused by the DoS attacks in \eqref{tau_p} with $\bar{\tau}_{i,k} \in [\tau_{\rm min}, \tau_{\rm max}]$. If there exist matrices
	$\mathcal{W}=\mathcal{W}^T>0, \mathcal{W}\in \mathbb{R}^{(3+p)\times (3+p)}$, $\mathcal{Y}\in \mathbb{R}^{3\times 3}$, $ \mathcal{Z}\in \mathbb{R}^{(3+p)\times (3+p)}$, and a scalar $0 < \mu <1$, such that
	\begin{align}
	\begin{bmatrix}
	\mathcal{Z}+\mathcal{Z}^T-\mathcal{W} &\mathcal{Z}^T\mathcal{S}_{M,j}^T-[\mathcal{Y}~~\textbf{\rm 0}]^T\mathcal{S}_{H,j}^T\\
	\star &(1-\mu )\mathcal{W}
	\end{bmatrix}>0,
	\end{align}
	where
	\begin{align}
	\mathcal{Z}=\begin{bmatrix}
	\mathcal{Z}_1 &\textbf{\rm 0}\\
	\mathcal{Z}_2 &\mathcal{Z}_3
	\end{bmatrix}\notag
	\end{align}
with $\mathcal{Z}_1\in \mathbb{R}^{3\times 3}$, $ \mathcal{S}_{M,j} \in \bar{\mathscr{M}}$, $\mathcal{S}_{H,j} \in \bar{\mathscr{H}}$, and $j=\{1,2,...,2^\nu\}$; then, the controller gain $K=\mathcal{Y}\mathcal{Z}_1^{-1}$ guarantees that the origin of platoon system \eqref{augumentplatoonmodel} is globally asymptotically stable.
\end{theo}

\textbf{\textit{Proof}:} See Appendix B.

\subsection{Controller Synthesis for String Stability}
To guarantee the string stability of the vehicle platoon under the DoS attacks, we explore the platoon control system synthesis method using the definition of $\mathcal{L}_2$ string stability.

Considering the driving terms from the preceding vehicle and the external disturbance, the discrete-space distributed platoon model in \eqref{augumentplatoonmodel} can be extended as
\begin{align}
\mathcal{X}_{i,k+1}=\bar{A}(\bar{\tau}_{i,k})\mathcal{X}_{i,k}+\bar{B}(\bar{\tau}_{i,k})\bar{u}_{i,k}+\mathcal{L}y_{i-1,k} +\mathcal{G}\bar{d}_{i,k} \label{stringplatoonmodel}
\end{align}
for $i, i\in \mathcal{V}_N$, with $\bar{A}(\bar{\tau}_{i,k})$ and $\bar{B}(\bar{\tau}_{i,k})$ in \eqref{augumentplatoonmodel}, and 
\begin{align}
\mathcal{L}:=\begin{bmatrix} 
\int_{0}^{h}e^{\bar{A}_0s}ds\bar{B}_2 \\\notag
0 \\\notag
\vdots \\\notag
0
\end{bmatrix},~
\mathcal{G}:=\begin{bmatrix} 
\int_{0}^{h}e^{\bar{A}_0s}ds\bar{B}_3 \\\notag
0\\\notag
\vdots \\\notag
0
\end{bmatrix}.
\end{align}

The following conditions guarantee the string stability of the platoon and attenuation of disturbances throughout the platoon.
\begin{theo}\label{theo2}
	Consider the discrete-space vehicular platoon model described in \eqref{stringplatoonmodel}.
	If there exist matrices
	$\tilde{\mathcal{W}}=\tilde{\mathcal{W}}^T>0, \tilde{\mathcal{W}}\in \mathbb{R}^{(3+p)\times (3+p)}$, $\tilde{\mathcal{Y}}\in \mathbb{R}^{3\times 3}$, $ \tilde{\mathcal{Z}}\in \mathbb{R}^{(3+p)\times (3+p)}$, and given scalars $0 < \sigma \leq 1$, $\gamma>0$, $a>0$, and $b>0$ satisfying the following LMIs 
    \begin{align}
	\begin{bmatrix}
		a(\tilde{\mathcal{Z}}^T+\tilde{\mathcal{Z}}-\tilde{\mathcal{W}}) &\tilde{\Omega}_j^T &\rm 0 &\tilde{\mathcal{Z}}^T{C}^T\\
	    \star &\tilde{\mathcal{W}} &\mathcal{L} &\rm 0\\
		\star &\star &b\sigma I &\rm 0\\
		\star &\star &\star &\frac{1}{b}I
	\end{bmatrix}
	\geq 0 ,\label{stringcon1}
    \end{align}
    \begin{align}
    \begin{bmatrix}
    	a(\tilde{\mathcal{Z}}^T+\tilde{\mathcal{Z}}-\tilde{\mathcal{W}}) &\tilde{\Omega}_j^T &\rm 0 &\tilde{\mathcal{Z}}^T{C}^T\\
    	\star &\tilde{\mathcal{W}} &\mathcal{G} &\rm 0\\
    	\star &\star &b\gamma I_2 &\rm 0\\
    	\star &\star &\star &\frac{1}{b}I
    \end{bmatrix}
    \geq 0 , \label{stringcon2}
    \end{align}
	where
	\begin{align}
	\tilde{\mathcal{Z}}=\begin{bmatrix}
	\tilde{\mathcal{Z}}_1 &\rm 0\\
	\tilde{\mathcal{Z}}_2 &\tilde{\mathcal{Z}}_3
	\end{bmatrix},\notag
	\end{align}
	$\tilde{\mathcal{Z}}_1\in \mathbb{R}^{3\times 3}$, $\tilde{\Omega}_j^T=\tilde{\mathcal{Z}}^T\mathcal{S}_{M,j}^T-[\tilde{\mathcal{Y}}^T ~\rm 0]^T\mathcal{S}_{H,j}^T$ with $ \mathcal{S}_{M,j} \in \bar{\mathscr{M}}$, $\mathcal{S}_{H,j} \in \bar{\mathscr{H}}$, and $j=\{1,2,...,2^\nu\}$. Then, the controller gain  $K=\tilde{\mathcal{Y}}\tilde{\mathcal{Z}}_1^{-1}$ guarantees $\mathcal{L}_2$ string stability, i.e., $||y_{i,k}||_{\mathcal{L}_2} \leq ||y_{i-1,k}||_{\mathcal{L}_2}$, 
	with a disturbance attenuation level $\gamma$ such that $||y_{i,k}||_{\mathcal{L}_2} \leq \gamma||d_{i,k}||_{\mathcal{L}_2}$.
\end{theo}

\textbf{\textit{Proof}:}See Appendix C.

\begin{rem}\label{remark6} 
	Some existing resilient platooning control approaches against DoS attacks, e.g., \cite{zhao2021resilient},\cite{zhang2020distributed},\cite{ge2022resilient}, \cite{xiao2021secure}, focus on presenting secure mechanisms to stabilize platoon control systems without considering string stability. 
	Another comparable study, e.g., \cite{merco2020hybrid}, designs the hybird controllers to resist DoS attacks in vehicular platoon with satisfying string stability.
	However, the presented method in \cite{merco2020hybrid} is conservative in robustness/resilience against DoS attacks, since the controller gains are chosen and fixed to find the maximum tolerated DoS attacks using the gridding search method, which also suffers heavy computation burden.
	In this paper, facilitated by the overapproximation techniques, we develop a resilient controller synthesis method that are maximally robust against DoS attacks and achieve internal and string stability.
\end{rem}
\section{Simulation results}
\begin{figure}
	\centering\includegraphics[height=5cm, width=8cm]{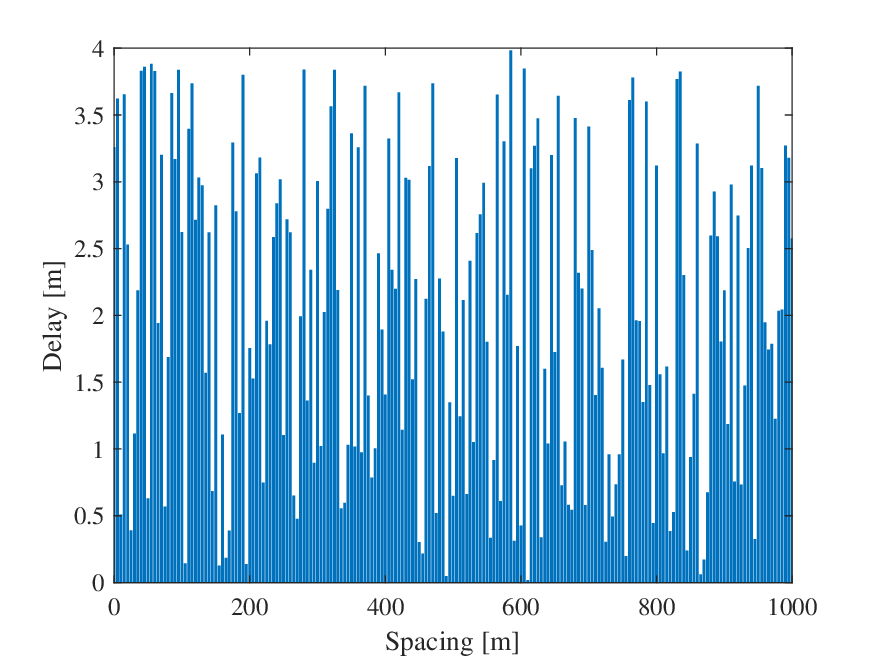}
	\centering\caption{Profile of the stochastic delays $\bar{\tau}_{i,k}$ induced by DoS attacks. \label{delay}}
\end{figure}

In this section, a vehicular platoon including eight vehicles (a lead vehicle and seven following vehicles) is considered in a numerical simulation. 
The main parameters of the spacing policy \eqref{spacingpolicy}, vehicle model \eqref{spacevecledynamic}, and LMI conditions in Theorem 2 are listed in Table \ref{parameter}. 
The sampling interval is set to $h=0.5 \text{m}$. The time gap of the desired spacing policy is set to $\Delta T=1 \text{s}$. 
The  time gap tracking error weights defined in \eqref{E1} are set as $\varepsilon_0=0.5$ and $\varepsilon=2$.
The simulation time interval is $[0, 1000~\text{m}]$ (the simulation is carried out in the space domain).
The external disturbances are described by $\bar{d}_i(s)=\text{sin}(0.01s), i \in \mathcal{V}_N^0$.
The reference velocity profile in the spatial domain is given by 
\begin{align}
v_\text{ref}(s)=\left\{\begin{aligned}
&20,~~~~s\in[0,100),
\\ &20+0.5(1-\text{cos}(0.01\pi(s-100))),s\in[100,300),
\\ &20,~~~~s\in[300,400),
\\ &20-0.6(1-\text{cos}(0.01\pi(s-400))),s\in[400,600),
\\ &20,~~~~s\in[600,1000],
\end{aligned}\notag
\right.
\end{align}
\begin{table}\renewcommand{\arraystretch}{1.35}
	\caption{Parameter Value Setting}\centering
	\label{parameter}
	\begin{tabular}{lccc}
		\toprule
		\hline
		Parameter &Value &Parameter &Value	\\
		\hline
		$h$   &0.5  &$\tau_{\text{min}}$   &0  \\
		$\Delta T$   &1  &$\tau_{\text{max}}$   &0.5 \\
		$\varepsilon$   &2  &$a$   &0.99  \\
		$\varepsilon_0$   &0.5  &$b$   &10  \\
		$\zeta$   &0.54  &$\sigma$   &0.8 \\
		\hline
		\hline
	\end{tabular}
\end{table}
The initial conditions are randomly generated.
The lower bound and upper bound of the functions $q_n(\bar{\tau}_{i,k})$ for the polytopic overapproximation in \eqref{upperlowerbound} are given as follows:
$\underline{q}_1(\bar{\tau}_{i,k})=0$, $\overline{q}_1(\bar{\tau}_{i,k})=0.5$, $\underline{q}_2(\bar{\tau}_{i,k})=0$,
$\overline{q}_2(\bar{\tau}_{i,k})=0.25$,
$\underline{q}_3(\bar{\tau}_{i,k})=0.3679$,
$\overline{q}_3(\bar{\tau}_{i,k})=1$.
By selecting $a=0.99$, $b=10$, and $\sigma=0.8$, and solving a set of LMIs \eqref{stringcon1} and \eqref{stringcon2} in Theorem 2, the controller gains can be obtained corresponding to the different upper bounds of the spatial delays induced by DoS attacks as summarized in Table \ref{controllergain}.
The parameter $p$ represents the maximum integer of multiples of sampling interval $h$ modeled in \eqref{tau_p}.
Note that when $p>7$, the controller gains are unavailable since feasible solutions cannot be found anymore. 
It means that the maximum spatial delay is eight times the sampling interval $h$, which can  theoretically be tolerated in the platoon system \eqref{augumentplatoonmodel}.
Therefore, in the simulation, the stochastic spatial delays are assumed to change as shown in Fig. \ref{delay}, under the effects of DoS attacks.
The maximum spatial delay is 4 m. 
Moreover, the controller gains are obtained using Theorem 2 as $K=[3\times10^{-6}, 0.0006, 0.0255]$ in the case of $p=7$.

\begin{table}\renewcommand{\arraystretch}{1.35}
	\caption{Controller gain}\centering
	\label{controllergain}
	\begin{tabular}{l|ccc}
		\toprule
		\hline
		Interger &$k_1$ &$k_2$ &$k_3$	\\
		\hline
  	   $p=1$   &0.0010  &0.2000   &0.1612  \\
  	   $p=3$   &0.0002  &0.0107   &0.1089  \\
  	   $p=5$   &0.0001  &0.0022   &0.0472  \\
  	   $p=7$   &$3\times10^{-6}$   &0.0006   &0.0255  \\
  	   $p>7$   &N/A  &N/A   &N/A  \\
		\hline
		\hline
	\end{tabular}
\end{table}

\begin{figure}
	\centering\includegraphics[height=6cm, width=9cm]{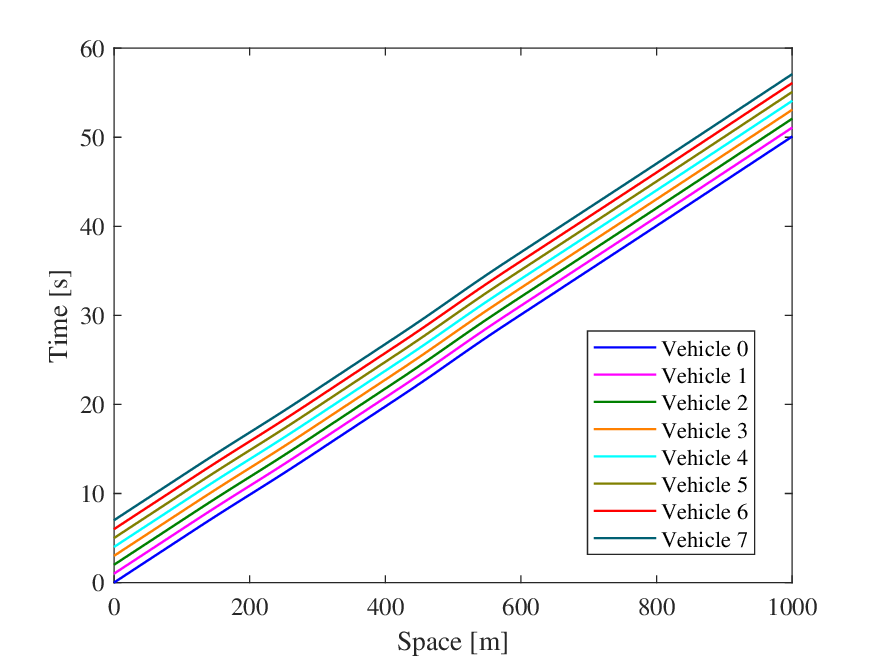}
	\centering\caption{Time-space trajectory profile for the vehicles in the platoon under DoS attacks. \label{position}}
\end{figure}

\begin{figure}
	\centering\includegraphics[height=6cm, width=9cm]{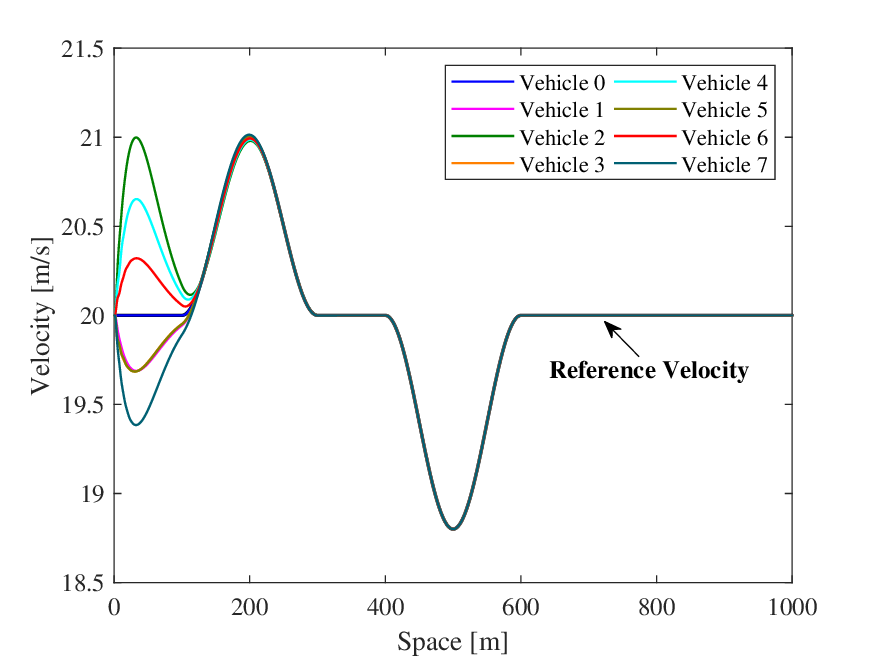}
	\centering\caption{Velocity profile for the vehicles in the platoon under DoS attacks. \label{velocity_nodis}}
\end{figure}

\begin{figure}
	\centering\includegraphics[height=6cm, width=9cm]{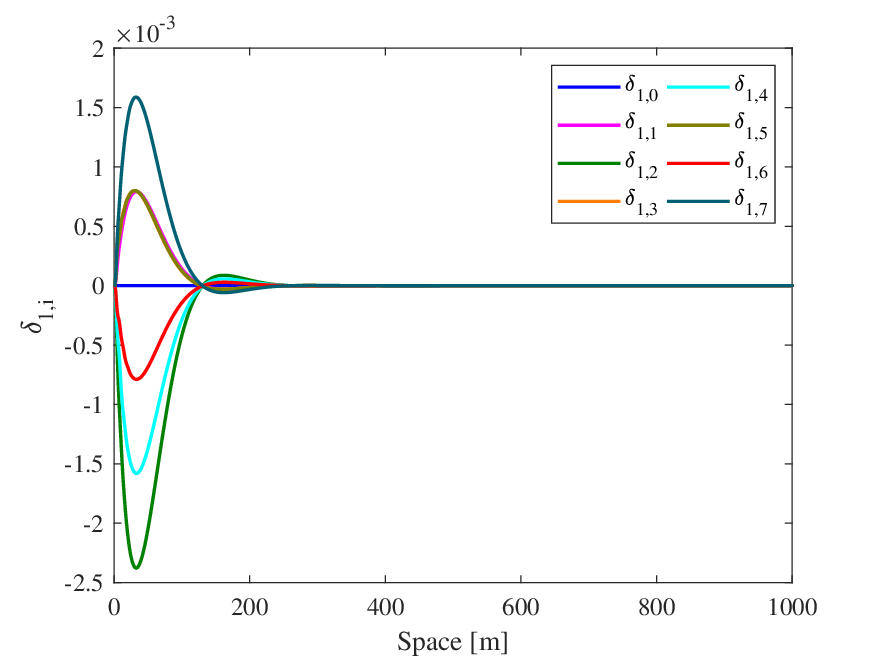}
	\centering\caption{Profile of the velocity tracking error $\delta_{1,i}$ as in \eqref{delta1} for the vehicles in the platoon under DoS attacks. \label{delta_1i_nodis}}
\end{figure}

\begin{figure}
	\centering\includegraphics[height=6cm, width=9cm]{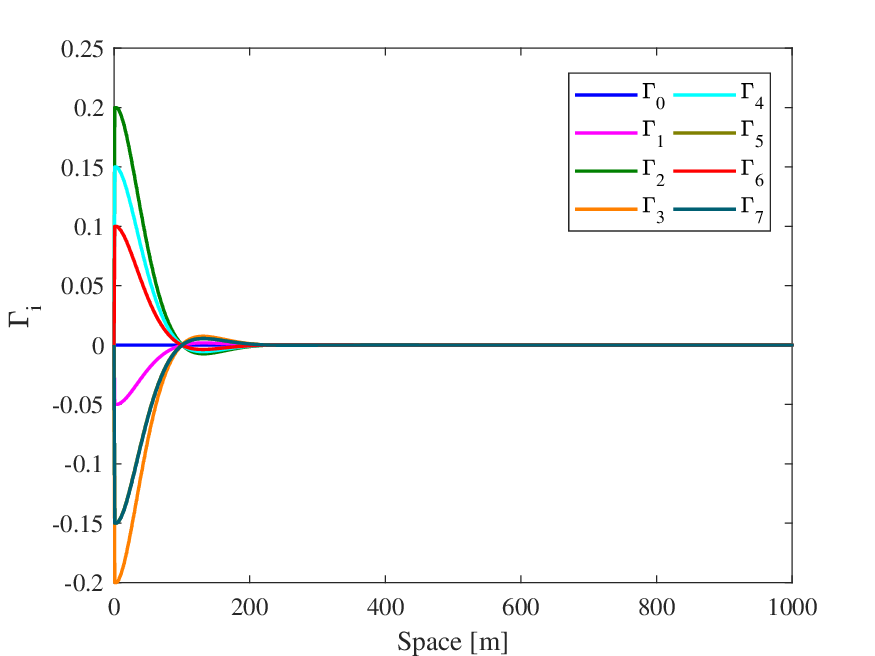}
	\centering\caption{Profile of the timing error $\Gamma_i$ as in \eqref{spacingpolicy} for the vehicles in the platoon under DoS attacks. \label{Gamma_nodis}}
\end{figure}

In the absence of external disturbances, the simulation results of the eight vehicles under DoS attacks are shown in Fig. \ref{position} to Fig. \ref{Gamma_nodis}.
Fig. \ref{position} shows the time-space trajectories of vehicles where a cohesive platoon is maintained with a desired headway.
Fig. \ref{velocity_nodis} shows that each vehicle can track the velocity profile imposed by reference velocity under the DoS attacks.
As shown in Fig. \ref{delta_1i_nodis}, the velocity tracking errors of the vehicles defined in \eqref{delta1} can converge to zero.
Fig. \ref{Gamma_nodis} shows the timing error profile of the vehicles,
which indicates that the desired spacing policy (\ref{spacingpolicy}) is satisfied.
From Fig. \ref{delta_1i_nodis} and Fig. \ref{Gamma_nodis}, 
one can conclude that the platoon control system is asymptotically stable, i.e., $\lim\limits_{s \to \infty} \Gamma_i(s)=0$ and $\lim\limits_{s \to \infty} \mathcal{E}_i(s)=0$, as defined in (\ref{E1}) and (\ref{E2}). Hence, the internal stability of the platoon system is achieved using the designed platoon controller and thereby guarantees the resilience of the platoon system under large DoS attacks.


\begin{figure}
	\centering\includegraphics[height=6cm, width=9cm]{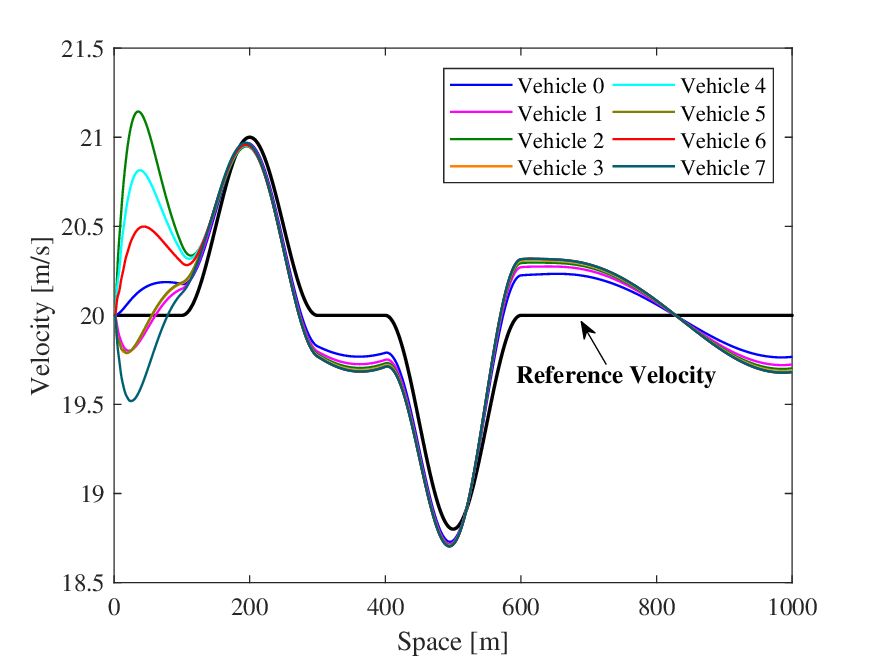}
	\centering\caption{Velocity profile of the vehicular platoon under DoS attacks and external disturbances. \label{velocity}}
\end{figure}

\begin{figure}
	\centering\includegraphics[height=6cm, width=9cm]{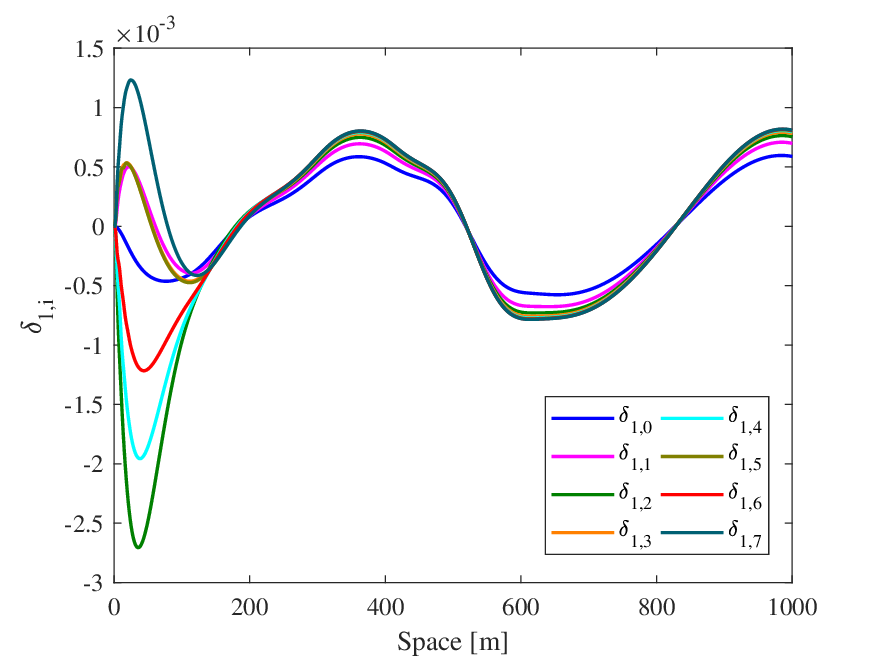}
	\centering\caption{Profile of the velocity tracking error $\delta_{1,i}$ as in \eqref{delta1} for the vehicles in the platoon under DoS attacks and external disturbances. \label{delta_1i}}
\end{figure}

\begin{figure}
	\centering\includegraphics[height=6cm, width=9cm]{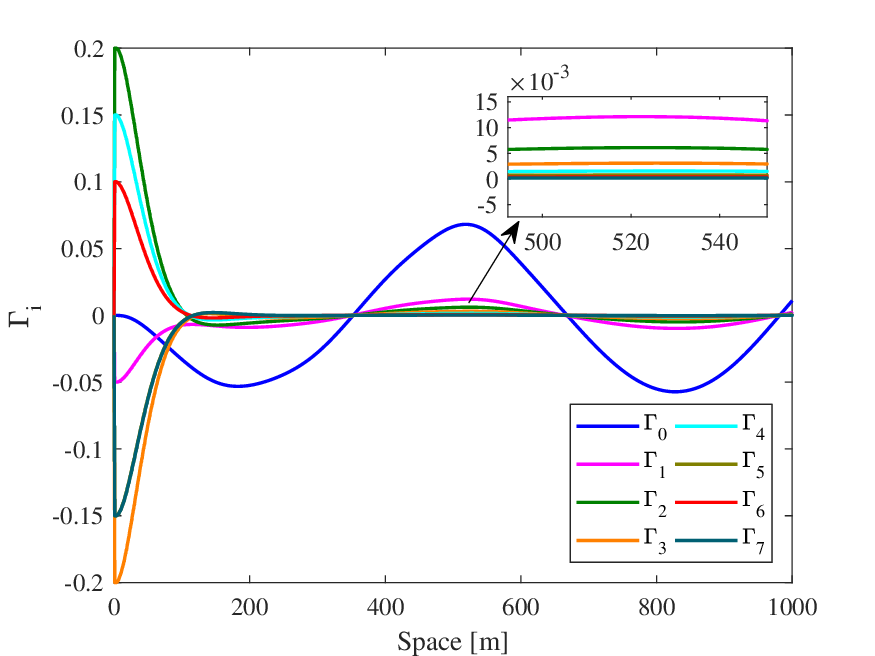}
	\centering\caption{Profile of the timing error $\Gamma_i$ as in \eqref{spacingpolicy} for the vehicles in the platoon under DoS attacks and external disturbances. \label{Gamma}}
\end{figure}

\begin{figure}
	\centering\includegraphics[height=6cm, width=9cm]{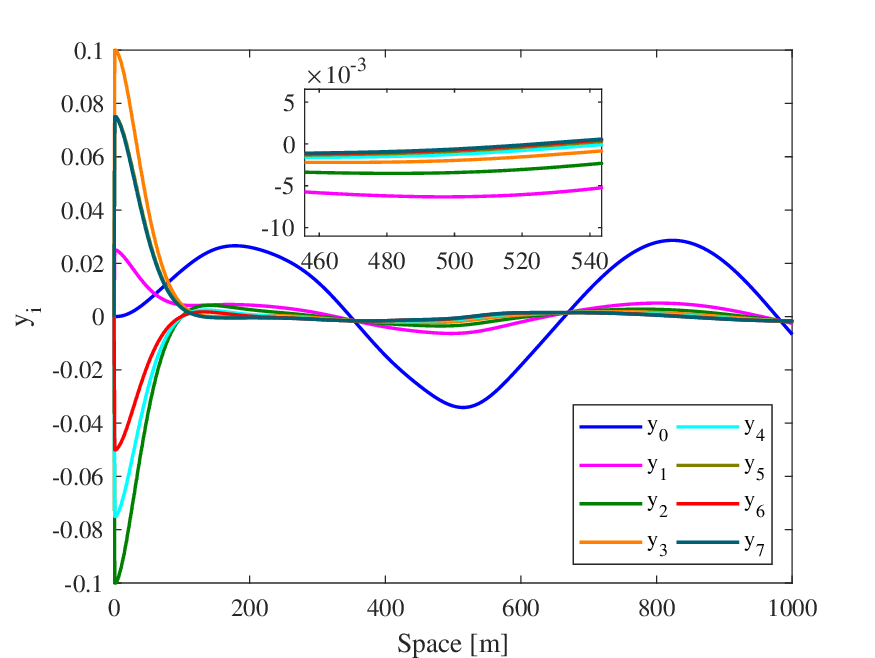}
	\centering\caption{Profile of the output of the vehicles in the platoon system under DoS attacks and external disturbances. \label{fig_output}}
\end{figure}

In the presence of both DoS attacks and external disturbance, the simulation results are exhibited in Fig. \ref{velocity} to Fig. \ref{output}.
Fig. \ref{velocity} shows the velocity profile of vehicles in the platoon, where it is clearly seen that each vehicle can track the prescribed reference velocity, although there exist fluctuations due to DoS attacks and external disturbances.
Fig. \ref{delta_1i} and Fig. \ref{Gamma} show the velocity tracking error and timing error profiles of the vehicles, respectively, which fluctuate around the equilibrium point under the influence of DoS attacks and external disturbances.
Besides, one can observe from the local zoom in Fig. \ref{Gamma} that $|\Gamma_0|>|\Gamma_1|>|\Gamma_2|>|\Gamma_3|>|\Gamma_4|>|\Gamma_5|>|\Gamma_6|>|\Gamma_7|$, which implies that the fluctuations are attenuated when propagating to the tail of the platoon.
Fig. \ref{fig_output} shows the output profile of the vehicles, where the string stability performance of the platoon system can be evaluated in terms of (\ref{stringstability}).
It is clearly seen from the local zoom in Fig. \ref{fig_output} that $|y_0|>|y_1|>|y_2|>|y_3|>|y_4|>|y_5|>|y_6|>|y_7|$, which indicates that the string stability of the platoon control system is guaranteed.
These simulation results validate the effectiveness of the proposed controller design method.


\begin{figure}
	\centering\includegraphics[height=6cm, width=9cm]{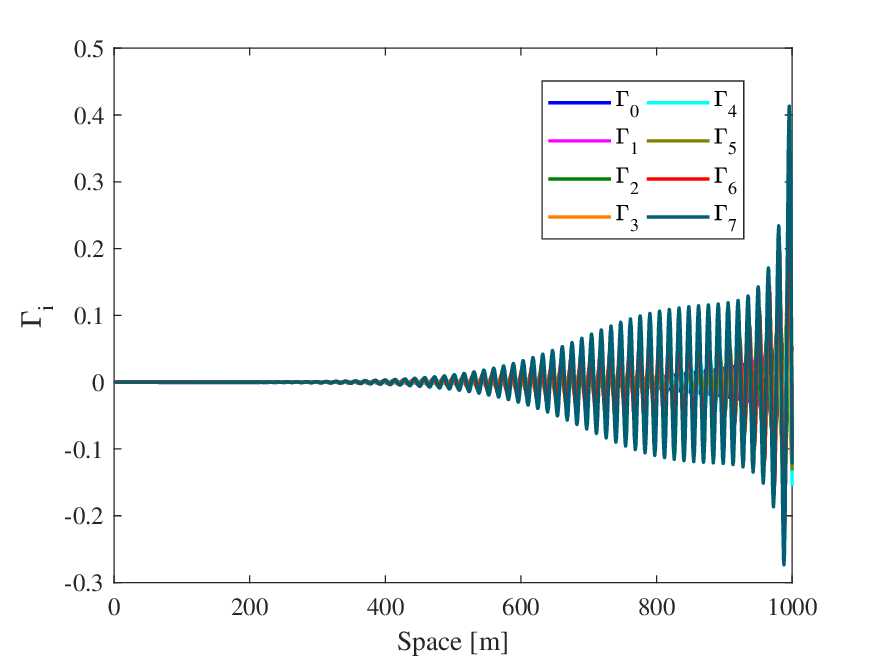}
	\centering\caption{Profile of the velocity tracking error $\Gamma_i$ under the controller presented in \cite{besselink2017string}. \label{compare_Gamma}}
\end{figure}

\begin{figure}
	\centering\includegraphics[height=6cm, width=9cm]{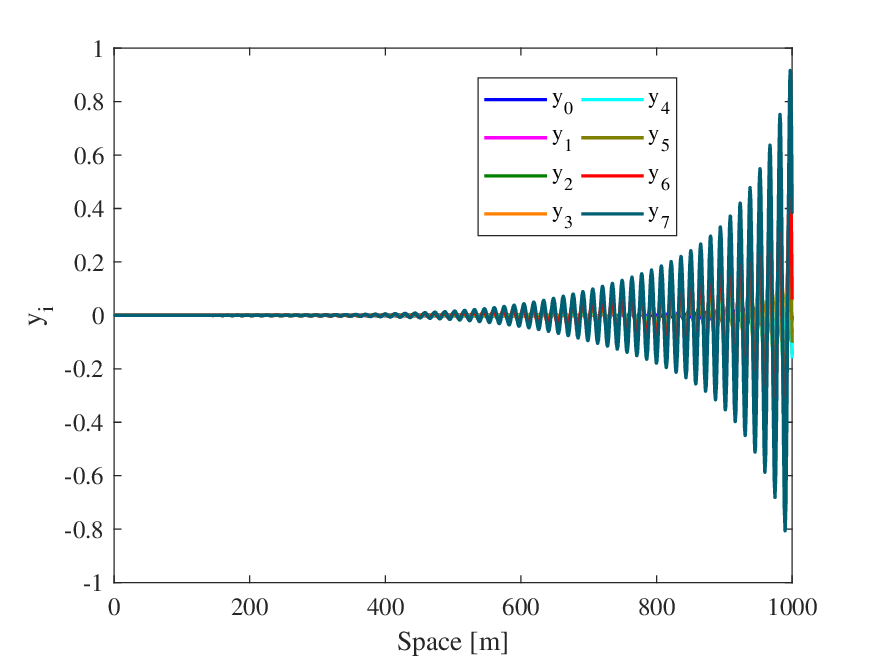}
	\centering\caption{Profile of the output of the vehicles in the platoon system under the controller presented in \cite{besselink2017string}. \label{com_output}}
\end{figure}

For comparison, the controller design method presented in \cite{besselink2017string} is evaluated under the same simulation environment, where the controller gains $K=[0, 0.09, 0.0025]$ as adopted in \cite{besselink2017string} are applied.
Fig. \ref{compare_Gamma} shows the velocity tracking error profile of the vehicles under the controller design method in \cite{besselink2017string}. 
In the presence of DoS attacks and external disturbances, it is evident that the internal stability of the platoon system cannot be guaranteed.
Likewise, the string stability of the platoon is also not available, as exhibited in Fig. \ref{com_output}.
By comparison, our controller design method is more robust under DoS attacks than the method presented in \cite{besselink2017string}.


\section{Conclusion}
In this paper, the platoon control problem has been addressed in the presence of DoS attacks and external disturbances.
A resilient controller synthesis framework that allows designing controllers for ensuring the internal stability and string stability of the platoon system in the spatial domain under long DoS attacks is presented.
The internal stability problem is formulated to guarantee the tracking of the desired spacing policy and the same prescribed reference velocity profile.
A distributed controller design approach is presented by means of polytopic overapproximations of the closed-loop dynamics under space delays induced by DoS attacks.
A set of matrix inequalities of sufficient conditions that guarantee string stability are derived and later used to find the controller gains that can tolerate the maximum spatial delay in the platoon system.
The simulation results reveal that the proposed control approach is effective and robust under DoS attacks and external disturbances.

Our future study will consider other types of cyber attacks and the limiting factors of the communication network in the platoon system.
Moreover, another future work is to explore other platoon control strategies with general communication topologies in the spatial domain.

\section*{Appendix}
\subsection{Proof of Lemma 2}
Based on the uncertain set $\mathscr{M}$ and $\mathscr{H}$  of the system matrices in the discrete-space model \eqref{polynomialplatoonmodel}, we obtain the overapproximation $\mathscr{M} \subseteq \bar{\mathscr{M}}$ and $\mathscr{H} \subseteq \bar{\mathscr{H}}$ in \eqref{F} and \eqref{G}, respectively. 
Note that $\bar{\mathscr{M}}$ and $\bar{\mathscr{H}}$ comprise $2^\nu$ vertices denoted by $\mathcal{S}_{M,j}$ and $\mathcal{S}_{H,j}, j=\{1,2,...,2^\nu\}$, respectively.
Moreover, any matrix in the set $\mathscr{M}$ or $\mathscr{H}$ can be represented as a convex combination of the generators of $\mathcal{S}_{M,j}$ or $\mathcal{S}_{H,j}$,
and is given by
\begin{align}
\begin{gathered}
\text{Co}(\bar{\mathscr{M}})=\left\{\sum_{j=1}^{2^\nu} \sigma_j\mathcal{S}_{M,j} |\sum_{j=1}^{2^\nu} \sigma_j=1, \sigma_j \in [0,1]                   \right\}, \\
\text{Co}(\bar{\mathscr{H}})=\left\{\sum_{j=1}^{2^\nu} \sigma_j\mathcal{S}_{H,j} |\sum_{j=1}^{2^\nu} \sigma_j=1, \sigma_j \in [0,1]                   \right\},
\end{gathered}
\end{align}
respectively, in the sense that
\begin{align}
\begin{gathered}
\mathscr{M} \subseteq \text{Co}(\bar{\mathscr{M}}), ~
\mathscr{H} \subseteq \text{Co}(\bar{\mathscr{H}}).           
\end{gathered}
\end{align}

Since \eqref{lemma2} holds for $\mathcal{S}_{M,j} \in \bar{\mathscr{M}}$, $\mathcal{S}_{H,j} \in \bar{\mathscr{H}}$, and $j=\{1,2,...,2^\nu\}$, for any $\sigma_j\geq 0$ and $\sum_{j=1}^{2^\nu}\sigma_j=1$, we have 
\begin{align}
\begin{gathered}
\sum_{j=1}^{2^\nu}\sigma_j\begin{bmatrix}
(1-\mu)P &(\mathcal{S}_{M,j}^T-\bar{K}^T\mathcal{S}_{H,j}^T)P\\
\star &P
\end{bmatrix}=\\
\begin{bmatrix}
(1-\mu)P &\sum_{j=1}^{2^\nu}\sigma_j(\mathcal{S}_{M,j}^T-\bar{K}^T\mathcal{S}_{H,j}^T)P\\
\star &P
\end{bmatrix}>0
\end{gathered}.
\end{align}

Therefore, it holds that
\begin{align}
\begin{bmatrix}
(1-\mu)P &(\bar{\mathcal{M}}^T-\bar{K}^T\bar{\mathcal{H}}^T)P\\
\star &P
\end{bmatrix}>0, \label{lem2}
\end{align}
$\forall \bar{\mathcal{M}} \in \text{Co}(\bar{\mathscr{M}})$ and $\forall \bar{\mathcal{H}} \in \text{Co}(\bar{\mathscr{H}})$. Using the Schur complement to \eqref{lem2}, the conditions \eqref{lem1} in Lemma 1 are obtained, which proves that the global asymptotic stability of the platoon system \eqref{augumentplatoonmodel} is guaranteed. $\square$

\subsection{Proof of Theorem 1}
According to Lemma 2, the platoon system \eqref{augumentplatoonmodel} is globally asymptotically stable if the following conditions of LMIs are satisfied
\begin{align}
\begin{bmatrix}
(1-\mu )P &(\mathcal{S}_{M,j}^T-\bar{K}^T\mathcal{S}_{H,j}^T)P\\
\star &P
\end{bmatrix}>0,
\end{align}
with $ \mathcal{S}_{M,j} \in \bar{\mathscr{M}}$ and $\mathcal{S}_{H,j} \in \bar{\mathscr{H}}$, and $j=\{1,2,...,2^\nu\}$. Let $P=\mathcal{W}^{-1}$, using the Schur complement, we have 
\begin{align}
\begin{bmatrix}
\mathcal{W}^{-1} &(\mathcal{S}_{M,j}^T-\bar{K}^T\mathcal{S}_{H,j}^T)\mathcal{W}^{-1}\\
\star &(1-\mu)\mathcal{W}^{-1}
\end{bmatrix}>0,
\end{align}
which is equivalent to	
\begin{align}
\begin{bmatrix}
\mathcal{Z}^T & \\
&\mathcal{W}
\end{bmatrix}
\begin{bmatrix}
\mathcal{W}^{-1} &\mathcal{D}_j^T\mathcal{W}^{-1}\\
\star &(1-\mu)\mathcal{W}^{-1}
\end{bmatrix}
\begin{bmatrix}
\mathcal{Z} & \\
&\mathcal{W}
\end{bmatrix}>0, \label{57}
\end{align}
where $\mathcal{D}_j=\mathcal{S}_{M,j}-\mathcal{S}_{H,j}\bar{K}$.
Then, satisfying \eqref{57} leads to
\begin{align}
\begin{bmatrix}
\mathcal{Z}^T\mathcal{W}^{-1}\mathcal{Z} &\mathcal{Z}^T\mathcal{D}_j^T\\
\star &(1-\mu)\mathcal{W} \label{66}
\end{bmatrix}
>0.
\end{align}

Since $\mathcal{Z}$ is of full rank and $\mathcal{W}$ is positive definite, we have 
\begin{align}
(\mathcal{W}-\mathcal{Z})^T\mathcal{W}^{-1}(\mathcal{W}-\mathcal{Z})\geq 0,
\end{align}
which is equivalent to
\begin{align}
\mathcal{Z}^T\mathcal{W}^{-1}\mathcal{Z}\geq \mathcal{Z}^T+\mathcal{Z}-\mathcal{W}. \label{lm3_ineq}
\end{align}

In terms of \eqref{66} and \eqref{lm3_ineq}, the following LMIs are equivalently satisfied
\begin{align}
\begin{bmatrix}
\mathcal{Z}^T+\mathcal{Z}-\mathcal{W} &\mathcal{Z}^T\mathcal{D}_j^T\\
\star &(1-\mu)\mathcal{W}
\end{bmatrix}
>0.
\end{align}

Using the fact that $\mathcal{Y}=K\mathcal{Z}_1$, we have
\begin{align}
\begin{bmatrix}
\mathcal{Y} &0
\end{bmatrix}=\bar{K}
\begin{bmatrix}
\mathcal{Z}_1 &0\\
\mathcal{Z}_2 &\mathcal{Z}_3
\end{bmatrix},
\end{align}
and thus the following LMIs are obtained
\begin{align}
\begin{bmatrix}
\mathcal{Z}+\mathcal{Z}^T-\mathcal{W} &\mathcal{Z}^T\mathcal{S}_{M,j}^T-[\mathcal{Y}~~0]^T\mathcal{S}_{H,j}^T\\
\star &(1-\mu)\mathcal{W}
\end{bmatrix}>0
\end{align}
with $ \mathcal{S}_{M,j} \in \bar{\mathscr{M}}$, $\mathcal{S}_{H,j} \in \bar{\mathscr{H}}$, and $j=\{1,2,...,2^\nu\}$. The proof of Theorem 1 is completed. $\square$

\subsection{Proof of Theorem 2}
According to the definition of $\mathcal{L}_2$ string stability, under the zero initial conditions, the platoon system is string stable if the following relationship is satisfied
\begin{equation}
||y_{i,k}||_{\mathcal{L}_2}\leq \sigma ||y_{i-1,k}||_{\mathcal{L}_2}
\end{equation}
with $0<\sigma \leq 1$. 

Select a Lyapunov functional candidate for the
discrete-space platoon model \eqref{stringplatoonmodel} as
\begin{equation}
V_{k}=\mathcal{X}_{i,k}^T\tilde{P}\mathcal{X}_{i,k}.
\end{equation}

Taking the derivative of $V_{k}$ along the trajectories of \eqref{stringplatoonmodel}, we have
\begin{align}
\begin{aligned}
\Delta V_k&=V_{k+1}-aV_{k} \\ 
&=\mathcal{X}_{i,k+1}^T\tilde{P}\mathcal{X}_{i,k+1}-a\mathcal{X}_{i,k}^T\tilde{P}\mathcal{X}_{i,k} \\ 
&\leq -by_{i,k}^Ty_{i,k}+b\sigma y_{i-1,k}^Ty_{i-1,k}
\end{aligned}\label{lyapunov}
\end{align}

Then, substituting the platoon dynamics \eqref{stringplatoonmodel} into \eqref{lyapunov}, and for facilitating analysis, the inequality \eqref{lyapunov} can be rewritten as follows
\begin{align}
\begin{bmatrix} 
\mathcal{X}_{i,k}^T \\
y_{i-1,k}^T
\end{bmatrix}^T
\Xi
\begin{bmatrix} 
\mathcal{X}_{i,k} \\
y_{i-1,k}
\end{bmatrix} \leq 0, \label{olmi}
\end{align}
where 
\begin{align}
\Xi=
\begin{bmatrix}
\mathcal{D}_j^T\tilde{P}\mathcal{D}_j-a\tilde{P}+bC^TC &\mathcal{D}_j^T\tilde{P}\mathcal{L} \\\notag
\star &\mathcal{L}^T\tilde{P}\mathcal{L}-b\sigma I
\end{bmatrix},
\end{align}
with $\mathcal{D}_j=\mathcal{S}_{M,j}-\mathcal{S}_{H,j}\bar{K}$, for $ \mathcal{S}_{M,j} \in \bar{\mathscr{M}}$, $\mathcal{S}_{H,j} \in \bar{\mathscr{H}}$, $j=\{1,2,...,2^\nu\}$.

If the LMIs \eqref{olmi} hold, using the Schur complement on $\Xi \leq 0$ leads to the following relationship
\begin{align}
\begin{bmatrix}
a\tilde{P}-bC^TC &0 &\mathcal{D}_j^T\tilde{P} \\
\star &b\sigma I &\mathcal{L}^T\tilde{P} \\
\star &\star &\tilde{P}
\end{bmatrix}
\geq 0 .\label{lmi1}
\end{align}

Using the elementary transformation of the matrix on the left  side of \eqref{lmi1}, let $\tilde{P}=\tilde{\mathcal{W}}^{-1}$, and the following LMIs are equivalently satisfied
\begin{align}
\begin{bmatrix}
a\tilde{\mathcal{W}}^{-1}-bC^TC &\mathcal{D}_j^T\tilde{\mathcal{W}}^{-1} &0\\
\star &\tilde{\mathcal{W}}^{-1} &\tilde{\mathcal{W}}^{-1}\mathcal{L}\\
\star &\star &b\sigma I
\end{bmatrix}
\geq 0 \label{lmi2}
\end{align}

Pre- and post-multiplying both the sides of \eqref{lmi2} by $\text{diag}(\tilde{\mathcal{Z}}^T, \tilde{\mathcal{W}}, I)$ and $\text{diag}(\tilde{\mathcal{Z}}, \tilde{\mathcal{W}}, I)$, respectively, we have\\
\begin{align}
\begin{bmatrix}
a\tilde{\mathcal{Z}}^T\tilde{\mathcal{W}}^{-1}\tilde{\mathcal{Z}}-b\tilde{\mathcal{Z}}^TC^TC\tilde{\mathcal{Z}} &\tilde{\mathcal{Z}}^T\mathcal{D}_j^T &0\\
\star &\tilde{\mathcal{W}}^{-1} &\mathcal{L}\\
\star &\star &b\sigma I
\end{bmatrix}
\geq 0 .\label{lmi3}
\end{align}

Similar to the proof of Theorem 1, using the fact $\tilde{\mathcal{Z}}^T\tilde{\mathcal{W}}^{-1}\tilde{\mathcal{Z}} \geq \tilde{\mathcal{Z}}^T+\tilde{\mathcal{Z}}-\tilde{\mathcal{W}}$ and the Schur complement, LMIs \eqref{lmi3} are satisfied if the following LMIs hold
\begin{align}
\begin{bmatrix}
a(\tilde{\mathcal{Z}}^T+\tilde{\mathcal{Z}}-\tilde{\mathcal{W}}) &\tilde{\mathcal{Z}}^T\mathcal{D}_j^T &0 &\tilde{\mathcal{Z}}^TC^T\\
\star &\tilde{\mathcal{W}} &\mathcal{L} &0\\
\star &\star &b\sigma I &0\\
\star &\star &\star &\frac{1}{b}I
\end{bmatrix}
\geq 0 . \label{lmi4}
\end{align}

Let $\tilde{\Omega}=\mathcal{D}_j\tilde{\mathcal{Z}}=(\mathcal{S}_{M,j}-\mathcal{S}_{H,j}[K ~0])\tilde{\mathcal{Z}}$. The LMIs \eqref{lmi4} lead to the condition \eqref{stringcon1} satisfied, where the controller gain is given by $K=\tilde{\mathcal{Y}}\tilde{\mathcal{Z}}_1^{-1}$.

The attenuation of disturbance propagation of the platoon system can be guaranteed for a given allowed bound $\gamma>0$ that satisfies
\begin{equation}
||y_{i,k}||_{\mathcal{L}_2}\leq \gamma ||\bar{d}_{i,k}||_{\mathcal{L}_2}.
\end{equation}

Similar to the proof of string stability conditions in \eqref{stringcon1}, the condition \eqref{stringcon2} can be correspondingly obtained. 
The proof of Theorem 2 is complete. $\square$

\section*{Acknowledgment}
The research leading to these results has received funding from the European Union’s Horizon Europe programme under grant agreement No 101069748 – SELFY project.

\bibliographystyle{IEEEtran}
\bibliography{references}

\end{document}